\documentclass[preprint2,tighten]{aastex6}
\pdfoutput=1 
\usepackage{amsmath,amstext}
\usepackage[T1]{fontenc}
\usepackage{apjfonts} 
\usepackage[figure,figure*]{hypcap}
\usepackage{color}

\newcommand{\unit}[1]{\ensuremath{\, \mathrm{#1}}}

\usepackage{hyperref}
\usepackage{breakurl}
\shorttitle{Towards Space-like Photometric Precision from the Ground with Beam-Shaping Diffusers}
\shortauthors{Stefansson et al.}

\begin{document}
\title{Towards Space-like Photometric Precision from the Ground with Beam-Shaping Diffusers}
\author{Gudmundur Stefansson\altaffilmark{1,2,3,4,5}}
\author{Suvrath Mahadevan\altaffilmark{1,2,3}}
\author{Leslie Hebb\altaffilmark{6}}
\author{John Wisniewski\altaffilmark{7}}
\author{Joseph Huehnerhoff\altaffilmark{8,9,10}}
\author{Brett Morris\altaffilmark{8}}
\author{Sam Halverson\altaffilmark{11,12}}
\author{Ming Zhao\altaffilmark{1,2,3,13}}
\author{Jason Wright\altaffilmark{1,2,3}}
\author{Joseph O'rourke\altaffilmark{14}}
\author{Heather Knutson\altaffilmark{14}}
\author{Suzanne Hawley\altaffilmark{8}}
\author{Shubham Kanodia\altaffilmark{1}}
\author{Yiting Li\altaffilmark{1}}
\author{Lea M. Z. Hagen\altaffilmark{1,15}}
\author{Leo J. Liu\altaffilmark{1,2,3}}
\author{Thomas Beatty\altaffilmark{1,2,3}}
\author{Chad Bender\altaffilmark{1,2,16}}
\author{Paul Robertson\altaffilmark{11,1,2,3}}
\author{Jack Dembicky\altaffilmark{9}}
\author{Candace Gray\altaffilmark{9}}
\author{William Ketzeback\altaffilmark{9}}
\author{Russet McMillan\altaffilmark{9}}
\author{Theodore Rudyk\altaffilmark{9}}

\email{gudmundur@psu.edu}
\altaffiltext{1}{Department of Astronomy \& Astrophysics, The Pennsylvania State University, 525 Davey Lab, University Park, PA 16802, USA}
\altaffiltext{2}{Center for Exoplanets \& Habitable Worlds, University Park, PA 16802, USA}
\altaffiltext{3}{Penn State Astrobiology Research Center, University Park, PA 16802, USA}
\altaffiltext{4}{NASA Earth and Space Science Fellow}
\altaffiltext{5}{Leifur Eiriksson Foundation Fellow}
\altaffiltext{6}{Department of Physics, Hobart and William Smith Colleges, 300 Pulteney Street, Geneva, NY, 14456, USA}
\altaffiltext{7}{Homer L. Dodge Department of Physics and Astronomy, University of Oklahoma, 440 W. Brooks Street, Norman, OK 73019, USA}
\altaffiltext{8}{Department of Astronomy, Box 351580, University of Washington, Seattle, WA 98195, USA}
\altaffiltext{9}{Apache Point Observatory, 2001 Apache Point Road, Sunspot, New Mexico, NM 88349, USA}
\altaffiltext{10}{Hindsight Imaging, Inc., 233 Harvard St. Suite 316, Brookline, MA 02446, USA}
\altaffiltext{11}{NASA Sagan Fellow}
\altaffiltext{12}{Department of Physics and Astronomy, University of Pennsylvania, Philadelphia, PA 19104, USA}
\altaffiltext{13}{The New York Times, 620 Eight Avenue, New York, NY 10018, USA}
\altaffiltext{14}{Division of Geological and Planetary Sciences, California Institute of Technology, Pasadena, CA 91125, USA}
\altaffiltext{15}{Institute for Gravitation and the Cosmos, The Pennsylvania State University, University Park, PA 16802, USA}
\altaffiltext{16}{Steward Observatory, University of Arizona, 933 N. Cherry Ave., Tucson, AZ 85719, USA}

\begin{abstract}
We demonstrate a path to hitherto unachievable differential photometric precisions from the ground, both in the optical and near-infrared (NIR), using custom-fabricated beam-shaping diffusers produced using specialized nanofabrication techniques. Such diffusers mold the focal plane image of a star into a broad and stable top-hat shape, minimizing photometric errors due to non-uniform pixel response, atmospheric seeing effects, imperfect guiding, and telescope-induced variable aberrations seen in defocusing. This PSF reshaping significantly increases the achievable dynamic range of our observations, increasing our observing efficiency and thus better averages over scintillation. Diffusers work in both collimated and converging beams. We present diffuser-assisted optical observations demonstrating $62^{+26}_{-16}$ppm precision in 30 minute bins on a nearby bright star 16-Cygni A (V=5.95) using the ARC 3.5m telescope---within a factor of $\sim$2 of \textit{Kepler}'s photometric precision on the same star. We also show a transit of WASP-85-Ab (V=11.2) and TRES-3b (V=12.4), where the residuals bin down to $180^{+66}_{-41}$ppm in 30 minute bins for WASP-85-Ab---a factor of $\sim$4 of the precision achieved by the \textit{K2} mission on this target---and to 101ppm for TRES-3b. In the NIR, where diffusers may provide even more significant improvements over the current state of the art, our preliminary tests have demonstrated $137^{+64}_{-36}$ppm precision for a $K_S =10.8$ star on the 200" Hale Telescope. These photometric precisions match or surpass the expected photometric precisions of TESS for the same magnitude range. This technology is inexpensive, scalable, easily adaptable, and can have an important and immediate impact on the observations of transits and secondary eclipses of exoplanets.
\end{abstract}

\section{Introduction}
Exoplanet science has seen an explosion in productivity over the past decade. The \textit{Kepler} spacecraft \citep{borucki2010} has detected over 3000 planet candidates \citep{burke2015}. However, many of the \textit{Kepler} stars are faint, and difficult to follow up with ground-based facilities. After the the failure of the second \textit{Kepler} reaction wheel, the repurposed \textit{Kepler} mission, \textit{K2}, has sampled a different population of host stars, namely more nearby and brighter stars, better suited for follow-up efforts from the ground. This has resulted in synergistic efforts from space and the ground, to rapidly confirm and verify new planet candidates \citep[e.g.][]{vanderburg2014,vanderburg2016,crossfield2015}.

The Transiting Exoplanet Survey Satellite (TESS) is scheduled for launch in 2018 \citep{Ricker2014}. TESS will survey the whole sky for transiting exoplanets around the nearest and brightest stars, and is expected to find thousands of Neptunes, and dozens of Earth-sized planets \citep{Ricker2014,sullivan2015}. However, the majority of TESS targets will only be observed for 26 days, with significantly larger observational coverage only at the North and South ecliptic poles. Therefore, most of the planet candidates will only have few transits observed, and will require timely ground-based follow-up to confirm their planetary nature.

As such, follow-up observations from the ground, both photometric and spectroscopic, will be crucial in maximizing the TESS yield \citep{plavchan2015}. Followup of promising TESS candidates rapidly after discovery will also enable the community to best use valuable James Webb Space Telescope (JWST) time for precise atmospheric characterization via transit spectroscopy \citep[e.g.,][]{cowan2015,benneke2017,batalha2017}, and determine how best to align efforts to study the full phase curves of exoplanets to characterize the thermal profiles of their atmospheres.

However, on the ground, telescopes have to contend with the deleterious effects of the atmosphere, including scintillation---the observed intensity variations (or "twinkling") of stars---, transparency variations, differential extinction, seeing and telescope guiding effects, which all limit the achievable photometric precision.

There have been successes in circumventing these problems to achieve high differential photometric precisions from the ground. While a detailed comparison  of diffuser assisted photometry with current state of the art is presented in Section 7, we briefly discuss some current techniques and the precision levels achieved with them here for context. 

To reach high precisions, modern detectors can be read out quickly with low read noise and images coadded to reach high signal strengths (e.g., \cite{kundurthy2013}, achieving 306 ppm/minute with a fast frame transfer CCD). Furthermore, narrow-band filters can be used to desensitize photometric measurements from water column density changes, telluric absorption variations, atmospheric emission line fluctuations, and to observe bright stars with modest and large-size telescopes (e.g., \cite{Colon2012}, achieving 455 ppm/minute with a 10meter telescope and a narrow band filter). Narrow band filters can also be used in conjunction with polarimetry in novel specialized instruments such as PEPPER (a Polarization Encoding differential Photometer and PolarimetER), to reach high-precision self-differential photometry on a single star without any reference stars \citep{potter2006}. Perhaps the most popular technique to reach high-precision photometry from the ground, is to defocus the telecope to spread the light over many pixels, decreasing sensitivity to individual pixels effects, increasing observing efficiency and allowing more light to be collected per integration. This has been done successfully by many groups \citep[e.g.,][]{southworth2009,mann2011,fukui2016,croll2011,zhao2014}, and excellent results have been reported using defocusing with conventional CCDs in the optical (e.g., \cite{southworth2009} achieved 434 and 385ppm/minute photometric precisions using a 3.58m telescope, and \citep{fukui2016} achieved 423 ppm/minute photometric precisions on a 1.88m telescope), and also using NIR detector arrays (\citep{croll2011} achieved 860ppm/minute using a 3.6m telescope, and \citep{zhao2014} achieved 3195ppm/minute using a 5m telescope. Although capable of yielding very high-precision photometry, defocusing the telescope can result in location-dependent aberrations in the Point Spread Function (PSF), and bright spots that vary with seeing \citep{southworth2009} that can saturate the detector. Defocusing can also affect guiding precision (which in turn degrades photometric precision) unless the guider has an independent focusing mechanism. Orthogonal-transfer CCDs \citep[e.g.,][]{tonry1997,howell2003,johnson2009} can be used to shape the PSF on the detector itself without needing to defocus, which has been shown by \cite{johnson2009} to demonstrate excellent photometric precisions of 539ppm/minute on a 2.2m telescope. Although this may potentially be more robust and repeatable than the defocusing method, this method requires custom orthogonal-transfer CCDs which are still not very common. We again refer the reader to Section 7, which further discusses these efforts, puts them in further context, and compares them to the precision levels achieved in this work with diffusers.

In this work, we present a new and inexpensive technology to reliably reach high photometric precisions on bright stars, even in suboptimal observing conditions. We use a custom beam-shaping diffuser, created using specialized nanofabrication techniques, to deterministically 'mold' the stellar image into a stable top-hat pattern. By using this diffuser, we minimize atmospheric effects without defocusing the telescope. Furthermore, by spreading the light over many pixels, we minimize flat-fielding errors, while simultaneously increasing observing efficiency, allowing us to observe bright stars reliably without saturating. This technology is versatile, offers broad-band compatibility, and is capable of stabilizing stellar PSFs with diffusers placed in either converging or collimated beams. While such diffusers have been briefly explored in the context of precision photometry for the upcoming CHaracterising ExOPlanet Satellite (CHEOPS) mission \citep{magrin2014}, they were not part of the final CHEOPS design. This work represents the first published results of detailed characterization, testing, and on-sky results using diffuser-assisted photometry. Specifically, in the optical, we present on-sky high-precision demonstrations on Penn State's PlaneWave CDK 24" telescope of $246_{-81}^{+176}$ppm in 30 minute bins. Also in the optical, we present diffuser-assisted observations performed on the Astrophysical Research Consortium (ARC) 3.5m Telescope at APO using the Astrophysical Research Consortium Telescope Imaging Camera (ARCTIC) \citep{huehnerhoff2016}, of 16 Cyg A, and the transits of WASP 85 A b and TRES-3b, demonstrating precisions of $62^{+26}_{-16}$ppm, $180^{+66}_{-41}$ppm and $\sim$101ppm in 30 minute bins, respectively. Lastly, we present high-precision photometry in the near-infrared (NIR) on the 200" Hale telescope at Palomar using the Wide-field Infrared Camera (WIRC) \citep{wilson2003}, with a precision of $137^{+64}_{-36}$ppm in 30 minute bins. Our optical observations on ARCTIC match or surpass the precisions that are expected of the TESS spacecraft \citep{Ricker2014,sullivan2015} for the same stellar magnitude range in the same binning timescale\footnote{Although our NIR precision of $137^{+64}_{-36}$ppm in 30 minutes on a $K_S = 10.8$ magnitude star is also better than the expected precision of TESS on a $I_C = 10.8$ magnitude star, comparing our WIRC NIR results to TESS is not completely analogous to comparing our optical results on ARCTIC to TESS, as we discuss further in Section \ref{sec:precisioncomparison}.}.

The 30 minute diffuser assisted photometric precision levels presented in this paper are now beginning to approach (and in some cases exceed) 80ppm---the transit depth of an Earth around a Sun-like star, even in the presence of scintillation noise. We stress that diffusers can be used to improve the precisions across different telescope apertures. However, we expect that the most significant precision gains beyond the precisions reported by us here, will most likely come from incorporating diffusers on the largest telescopes, such as on the upcoming HiPERCAM on the 10m Gran Telescopio Canarias (GTC) or the new OCTOCAM instrument \citep{postigo2016} for the 8m Gemini telescopes, or on telescopes equipped with conjugate plane photometers to correct for scintillation \citep{osborn2011}.

This paper is structured as follows. Section 2 discusses issues and mitigation strategies in achieving high-precision photometry from the ground, setting the stage for the utility of diffusers. Section 3 gives a description of diffusers and how they can be used in telescopes for precision photometry applications. In Section 4, we describe our lab test setup, and our observations in the optical and the NIR. We present our lab and on-sky results in Section 5. Section 6 discusses our MCMC modeling and fits for the WASP 85 A b and TRES-3b transits. Section 7 provides further discussion and remarks on this technology and is applicability for use on other telescopes. We conclude in Section 8 with a summary of our findings.

\section{Reaching High Photometric Precisions from the Ground}
The empirical differential photometric precision achieved from telescopes in space and on the ground is well described by the theoretical calculation of noise for a well-behaved CCD \citep{merline1995}. 
Similar to the formalism outlined in \cite{collins2017}, the total photometric noise $N$ (excluding scintillation) in ADU (analog-to-digital unit) for a CCD aperture photometry measurement is

\begin{equation}
N = \frac{\sqrt{V_* + n_{\mathrm{pix}} \left(1 + \frac{n_{\mathrm{pix}}}{n_b} \right) \left(V_S + V_D + V_R + V_f \right)}}{G},
\label{eq:phot}
\end{equation}
 
where $G$ is the gain of the CCD in electrons/ADU, $V_*$ is the variance of the net background subtracted counts in the aperture from the star (unit: electrons$^2$), $n_{\textrm{pix}}$ is the number of pixels in the aperture, $n_b$ is the number of pixels used to estimate the mean background sky signal, $V_S$ is the variance in the sky background signal per pixel (unit: electrons$^2$/pixel), $V_D$ is the variance in the dark current signal per pixel (unit: electrons$^2$/pixel), and $V_R$ is the variance of the read noise per pixel (unit: electrons$^2$/pixel), and the last term $V_f$ is the variance in the digitization noise within the A/D converter (unit: electrons$^2$/pixel). Table \ref{tab:variances} lists the absolute values of the variances in Equation \ref{eq:phot}, relating them to the corresponding fluxes measured in number of ADUs or electrons as outlined in \cite{merline1995}, and \cite{collins2017}.

\begin{table}[t]
	\centering
	\caption{Absolute values of variances in Equation \ref{eq:phot} along with the underlying distribution of the variables. As presented in \cite{collins2017}, $F_*$ is the net background subtracted counts in the aperture from the star in ADUs, $F_S$ is the sky background signal in ADU/pixel, $F_D$ is the dark current signal in electrons/pixel, and $F_R$ is the read noise in electrons/pixel/read, and $\sigma_f$ is an estimate of the 1-$\sigma$ error introduced within the A/D converter with a value of $\sim$0.289ADU \citep{merline1995,collins2017}.}
	\begin{tabular}{l l l}
	\hline\hline
	  Variance & Distribution & Absolute value   \\ \hline
      $V_*$    & Poisson      & $|G F_*|$        \\ 
      $V_S$    & Poisson      & $|G F_S|$        \\ 
      $V_D$    & Poisson      & $|F_D|$          \\ 
      $V_R$    & Gaussian     & $|F_R^2|$        \\ 
      $V_f$    & Uniform      & $|G^2\sigma_f^2|$ \\
	\hline
	\end{tabular}
	\label{tab:variances}
\end{table}

Using a similar formalism as in \cite{collins2017}, the final normalized relative flux error on the relative flux $F_{\textrm{rel flux}} = F_T/F_E$ (where $F_T$ is the flux from the target star in ADU, and $F_E$ is the total integrated flux from the ensemble in ADU) is calculated using an ensemble of reference stars, using
\begin{equation}
\sigma_{\textrm{rel flux}} = F_{\textrm{rel flux}} \cdot \sqrt{\frac{N_T^2}{F_T^2} + \frac{N_E^2}{F_E^2}},
\label{eq:stdflux}
\end{equation}
where $N_T$ and $N_E$ are the noise from the target and the ensemble in ADU, respectively. 
For each individual star, the noise is calculated using Equation \ref{eq:phot}, and for the stars in the ensemble, the noise $N_E$ is the total noise from all of the reference stars added in quadrature. 
In normalized units, the corresponding relative flux error is given by,

\begin{equation}
\frac{\sigma_{\textrm{rel flux}} }{F_{\textrm{rel flux}}} = \sqrt{\frac{N_T^2}{F_T^2} + \frac{N_E^2}{F_E^2}}.
\label{eq:stdflux2}
\end{equation}

Hereafter we will refer to $\sigma_{\textrm{rel flux}}$ in these normalized units.

It is instructive to study Equations \ref{eq:phot}, and \ref{eq:stdflux2} in the special case where we assume minimal noise other than photon noise (i.e., near-perfect detector and minimal background sky noise) and a detector gain close to $G\sim1$. In this case the total noise in ADU is simply $N \sim \sqrt{F}$. Extending this to observations of a target star with numerous bright nearby reference stars (i.e., where the photon-noise contribution from the reference ensemble is minimal: $F_E \sim 0$), then we see from Equation \ref{eq:stdflux2} that,
\begin{equation}
\sigma_{\textrm{rel flux}} = \sqrt{\frac{N_T^2}{F_T^2}} \sim \sqrt{\frac{F_T}{F_T^2}} \sim \frac{1}{\sqrt{F_T}} \sim \frac{1}{N_T} ,
\label{eq:stdflux3}
\end{equation}
i.e., the normalized relative flux error also reduces to a similar square-root dependence with the target flux in ADUs.

Photometric errors are also introduced by instrumental effects which are not included in these equations, such as inhomogeneous pixel response of detectors and sensitivity to pixel position. Light curves are often decorrelated in some manner with respect to instrumental parameters in order to remove these effects from both space-based and ground-based photometry. Photometric errors are also introduced by limited electron well depth of detectors limiting integration times before saturation, which is problematic especially for larger telescopes observing bright stars such as the TESS stars.

From the ground, the precision of ground-based telescopes is further limited due to the deleterious effects of the Earth's atmosphere. Transparency fluctuations, sky background noise, scintillation and differential extinction can all affect photometric precision. In particular, scintillation due to turbulence in the upper atmosphere, is a particularly insidious source of photometric error for ground-based photometry \citep{young1967,dravins1998,kornilov2012,osborn2015}. Here we discuss the error sources particular to ground-based photometry and our mitigation strategies using diffusers.

\subsection{Atmospheric Noise Sources}
\subsubsection{Transparency variations} 
Transparency variations include shifting cloud cover. The exact variations will depend on the weather, and the observing site, and as such this source of noise is particularly difficult to estimate for any given observation. This is generally minimized through differential photometry, where the core assumption is that the transparency variations affect the target and reference stars equally. Transparency variations are further minimized by performing observations from a good observing site at high elevation. Efforts have been made to estimate the impact of this effect for some observing sites \citep[e.g.,][]{mann2011} using different atmospheric models, demonstrating that the median noise due to transparency variations is typically smaller than Poisson and scintillation noise for photometric nights \citep{mann2011}. For our observations, we assume that for a clear photometric night at a good observing site, this error source is typically much smaller than the expected Poisson and scintillation noise.

\subsubsection{Molecular absorption and differential extinction}
A related issue to transparency variations is variable molecular absorption. Commonly used broad-band filters, such as the SDSS $u^\prime g^\prime r^\prime i^\prime z^\prime$ \citep{fukugita1996} and UBVRI Johnson-Cousins filters \citep{bessell1990}, each operate over a wide bandpass and include a number of molecular absorption lines (with water, oxygen, and ozone being the primary absorbers in the optical). For water, the depth of these lines is strongly dependent on the water column at the time of observation, which is dependent on the exact weather conditions and the airmass of the target being observed. This effect is minimized by observing in a bandpass not contaminated by such lines.

Differential extinction is of two types: first order and second order. First-order differential extinction is caused by the variation in airmass difference of the target and the reference star throughout the observation, resulting in a relative brightness change. \cite{mann2011} estimated that the magnitude of this effect at Maunakea can be on the order of $\sim$$10^{-4}-10^{-3}$, depending on the observing conditions and passband being used. However, being a systematic trend correlated with airmass, \cite{mann2011} mention that this effect can generally be detrended out at high-precision if the extinction variation is minimal. Second-order differential extinction is caused by the target and the reference star not being of the same spectral type. Stars with different spectral types will vary differently with extinction throughout the observation. This color effect is smaller in redder passbands, with stars of later spectral type and narrower bandwidth filters, and can further be minimized by a judicious choice of reference stars of the same or similar spectral types.

\begin{figure}[t]
	\begin{center}
		\includegraphics[width=\columnwidth]{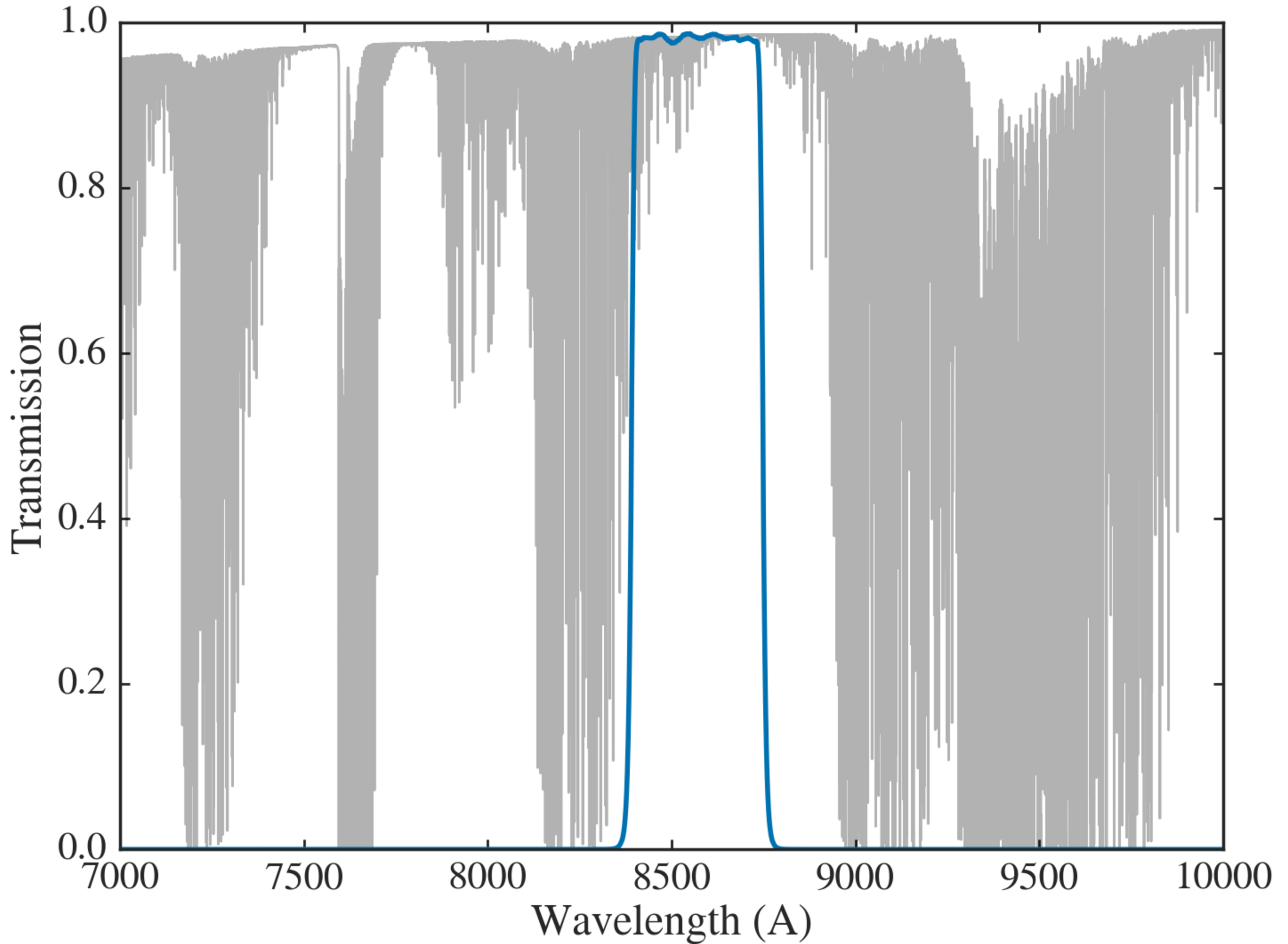}
	\end{center}
	\caption{Semrock filter transmission shown in blue. Shown in grey is atmospheric transmission calculated using TERRASPEC \citep{bender2012}. Filter transmission curve from the Semrock website (see text for details).}
	\label{fig:semrock}
\end{figure}

Both molecular absorption and extinction can be minimized by observing in a red-optical bandpass filter with little to no molecular absorption lines \citep{mann2011}. We explored the parameter space of commercially available filters, and converged on an off-the-shelf filter from Semrock (part number: 857/30), operating in a red passband between 842-872nm\footnote{\url{https://www.semrock.com/FilterDetails.aspx?id=FF01-857/30-25}}. Figure \ref{fig:semrock} shows the transmission curve of this filter, along with typical molecular absorption bands as calculated by TERRASPEC \citep{bender2012} around this region. As this filter is centered at the red-end of the optical spectrum, the recorded photometric signal will be less sensitive to variable Rayleigh scattering.

Thus, with an informed choice of bandpass filter in the red-optical, minimally contaminated by molecular absorption lines, we assume that scintillation and photon noise as the dominant error sources.

\subsubsection{Scintillation}
For bright nearby stars, the photometric precision is often not limited by photon noise or by background sky counts, but rather intensity fluctuations---or scintillation---produced by Earth's atmosphere \citep{osborn2015}. Scintillation is caused by the spatial intensity fluctuations crossing the pupil boundary, and the time-scale is determined by the wind speed of the turbulent layer \citep{young1967,dravins1998,osborn2011}.

The expected scintillation noise for a given star is described by \cite{young1967}, and \cite{dravins1998} in units of relative flux, with the following approximation:

\begin{equation}
\sigma_{s} = 0.09 D^{-\frac{2}{3}} \chi^{1.75} {\left(2t_{\mathrm{int}}\right)}^{-\frac{1}{2}} e^{\frac{-h}{h_0}},
\label{eq:stds2}
\end{equation}

where $D$ is the diameter of the telescope in centimeters, $\chi$ is the airmass of the observation, $t_{int}$ is the exposure time in seconds, and $h$ is the altitude of the telescope in meters, and $h_0\simeq8000\unit{m}$ is the atmospheric scale height. The constant 0.09 factor in front has a unit of $\unit{cm^{2/3}s^{1/2}}$, to give the scintillation error in units of relative flux.
This equation is approximate and highly reliant on the site and the strength and direction of winds in the upper atmosphere, and the exponent above the airmass term can range from 1.5 to 2.0 depending on the wind direction \citep{southworth2009,osborn2011}. However, for exposures longer than 1 second (long exposure regime for scintillation) the wind profile tends to average out \citep{kornilov2012,osborn2015}. Additionally, it has been suggested by \citet{osborn2015} that the median value of scintillation is a factor of 1.5 higher than suggested by Equation \ref{eq:stds2}. In the case of differential photometry, the strength of scintillation depends on the number of uncorrelated reference stars $n_E$ \citep{kornilov2012}. The degree of correlation depends on the angular separations of the stars from each other \citep{kornilov2012}, where 20" is generally the radius within which they are correlated. Combining these two terms, and assuming our target and reference stars are uncorrelated, we have the following equation for the scintillation for differential photometry,

\begin{equation}
\sigma_{\mathrm{scint}} = 1.5 \sigma_{s} \sqrt{1 + 1/n_{\mathrm{E}}},
\label{eq:stdscint}
\end{equation}

assuming $n_E$ uncorrelated reference stars. This illustrates the advantages of using multiple reference stars for precision photometry. The total error including scintillation is then,
\begin{equation}
\sigma_{\mathrm{tot}} = \sqrt{ \sigma_{\textrm{rel flux}}^2 + \sigma_{\mathrm{scint}}^2},
\label{eq:stdtot}
\end{equation}
assuming the other errors, e.g., from transparency variations and differential extinction are minimal.

\begin{figure}
	\begin{center}
		\includegraphics[width=\columnwidth]{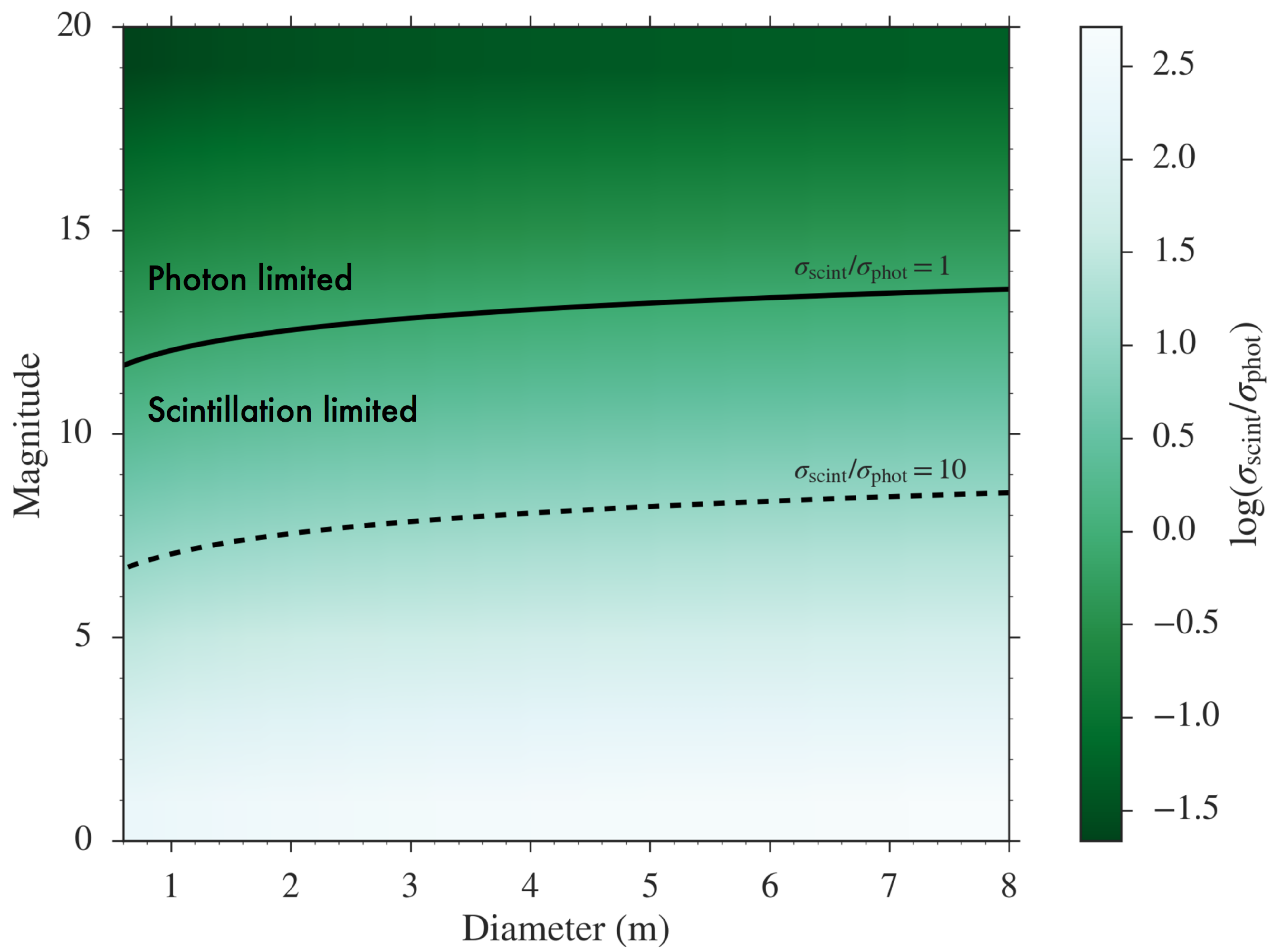}
	\end{center}
	\caption{Ratio of scintillation noise to photometric noise in the magnitude-telescope-diameter plane, assuming a perfect telescope (100\% throughput) and airmass of 1. The solid curve shows where the scintillation noise equals the shot noise. Therefore, stars below this line are scintillation limited, and above it photon limited. The dashed curve shows where the scintillation noise is an order of magnitude larger than the shot noise. Figure adapted from \cite{osborn2015}.}
	\label{fig:scint}
\end{figure}

As the photon and scintillation errors tend to be the largest sources of noise in ground-based photometry, it is instructive to look at the ratio of the two, $\sigma_{\mathrm{scint}}/\sigma_{\mathrm{phot}}$, to see when each dominates. To do this, we adapt a similar calculation and methodology as described by \citep{osborn2015}, showing the dependence of this ratio in the target star magnitude, and telescope diameter plane (Figure \ref{fig:scint}). The pure photon noise in normalized relative flux units $\sigma_{\textrm{phot}} = 1/N $, is calculated using Equation \ref{eq:phot} (see also Equation \ref{eq:stdflux3}), assuming no sky background, and a perfect detector (no read or dark noise) on a telescope with 100\% throughput, and the scintillation error is calculated using Equation \ref{eq:stdscint}, assuming an airmass of 1.0, an altitude of 2700m, and one reference star. For other observational parameters the results must be scaled accordingly. The solid black line in Figure \ref{fig:scint} shows where the scintillation and shot noise errors are equal. Therefore, stars below this curve ($\sigma_{\mathrm{scint}}/\sigma_{\mathrm{phot}} > 1$), are scintillation limited, and stars above this curve ($\sigma_{\mathrm{scint}}/\sigma_{\mathrm{phot}} < 1$) are photon limited. The dotted curve shows where the scintillation error is an order of magnitude larger than the photon noise ($\sigma_{\mathrm{scint}}/\sigma_{\mathrm{phot}} = 10$). The dependence of this ratio with diameter is $\sim$$D^{1/3}$, so the ratio will increase modestly with telescope aperture, but we stress that both error terms decrease with telescope aperture: scintillation, and photon noise, as $\sim$$D^{-2/3}$, and $\sim$$D^{-1}$, respectively. Due to this modest dependence with telescope diameter, as mentioned by \cite{osborn2015}, we can say that stars brighter than V-band mag of $\sim$13 will be scintillation limited across different telescopes\footnote{Similar to \citep{osborn2015}, we only plot the telescope diameter up to $8\unit{m}$, as the scintillation relation in Equation \ref{eq:stds2} is most accurate for this range of telescope diameters.}.

\subsection{Reaching high photometric precisions}
To reach high photometric precision requires that we consider all of the parameters discussed above. Our overall strategy follows many of the CCD photometry practices common in the field (see e.g., discussion by \cite{mann2011}). As mentioned by \citet{mann2011}, signals of $\sim$$10^7$ or higher are needed to achieve sub-millimag precisions. Spreading the PSF over a large number of pixels is a well established observing technique to reach such signal levels in a singe exposure. Spreading out the light increases exposure times before saturation, while simultaneously reducing scintillation errors and flat field errors due to variations in inter-pixel sensitivities. However, spreading out the light over many pixels increases background noise, which can be the dominant noise source on faint targets, and targets observed at redder wavelengths where the background sky is brighter.

\begin{figure}
	\begin{center}
		\includegraphics[width=\columnwidth]{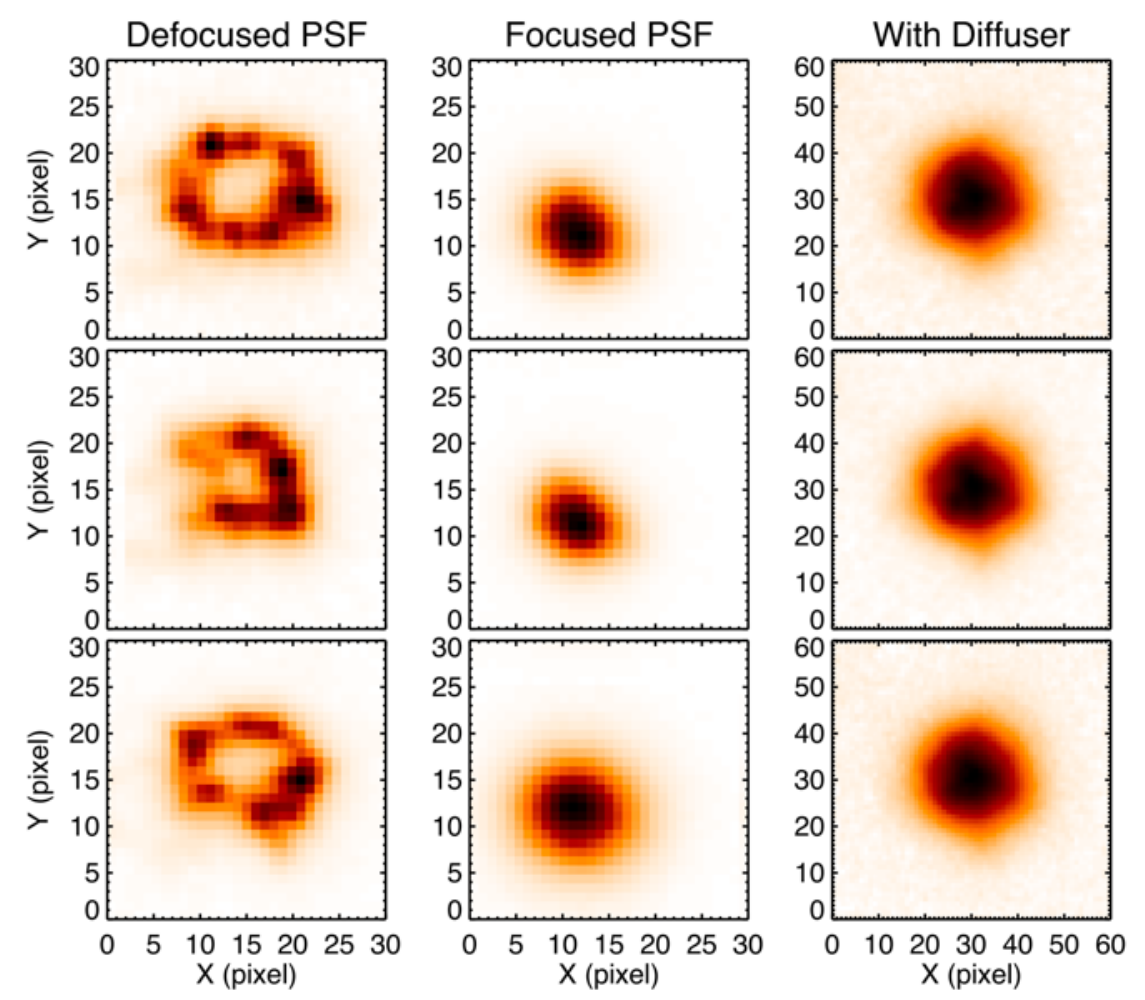}
	\end{center}
	\caption{Comparison of Palomar/WIRC PSFs under three observing modes at different epochs: defocused, focused, and diffused. The defocused PSF shows bright spots due to astigmatism of the telescope, inducing a significant amount of "red noise" to the light curve. Both defocused and focused modes show varying PSFs due to seeing variations, while the diffused PSF stays stable in shape (flux level still varies due to telluric fluctuations). Images with the diffuser shown at a different scale for clarity. A video version of this figure can be found in the online version of the manuscript.}
	\label{fig:defocusvsdiffuser}
\end{figure}

This has successfully been done on the ground by telescope defocusing \citep[e.g.,][]{southworth2009}, where the PSF is spread over many pixels through imaging the telescope pupil. However, defocusing often results in a "doughnut"-shaped PSF that is location-dependent across the imaging array, revealing numerous other optical aberrations \citep{howell2003}. In particular, a defocused image is subject to atmosphere-induced phase errors (seeing) in ways that an in-focus image is not (Figure \ref{fig:defocusvsdiffuser}). In both cases, phase errors from seeing will induce fluctuations across the face of the image, and even if the total flux may be conserved, the flux will be redistributed between pixels, producing uncertainties to the extent that the pixel responses are not perfectly calibrated \citep{mann2011}. For defocused images these phase-induced errors can create uneven signal distributions across the PSF, often resulting in time-varying high-intensity spikes (column 1 in Figure \ref{fig:defocusvsdiffuser}). These spikes cause the PSF to be asymmetric, induces more pixel-dependent errors, and can cause the detector to saturate. The time-varying PSF asymmetry can also cause the centroid of the PSF to shift, reducing photometric precision. For focused observations, the same seeing effect is present, but in this case, the phase errors are more localized around the center of the PSF (instead of being spread out over a doughnut), resulting in a broadened and blurred PSF instead (Figure \ref{fig:defocusvsdiffuser}). Despite these drawbacks of defocused images, defocused observations generally yield better photometric precisions than in-focused observations, due to the low dynamic range and high susceptibility to guiding and flat-field errors for in-focused observations.

An "in-focus" \textit{diffused} image brings out the best from both of these methods: allowing for a high dynamic range and  minimal flat-field and guiding errors, while minimizing any phase-induced errors due to seeing. This is illustratd in Figure \ref{fig:defocusvsdiffuser}, which compares a) the defocused and b) focused PSFs of WIRC to c) diffused-assisted observations with WIRC (where a "focused" image is deterministically spread out over many pixels). From Figure \ref{fig:defocusvsdiffuser}, we see that the defocused PSF changes significantly due to seeing variations (more so than the focused observations which vary and blur as well), while the diffused PSF is broad and stable throughout the observations. In particular, the defocused PSF shows numerous peaks whose locations and intensities change with time. A video version of this figure can be found in the online version of the manuscript.

\section{Description of Diffusers}
"Diffusers" is a generic term encompassing optical components or materials that use microscopic surface or bulk structures to control, shape and homogenize the distribution of light. Through precisely controlling the size, shape, location, and distribution of the surface structures used, an input beam can be molded to produce a desired output pattern with broad-band compatibility. Diffusers have a plurality of applications, including, but not limited to, use in the telecommunication industry, automotive, and architectural lighting applications. In this paper, we show that diffusers are also applicable for use in precise photometry.

\subsection{Diffuser Types}
Below we give an overview of four basic types of diffusers: ground-glass, holographic, diffractive, and Engineered Diffusers\texttrademark. 

\subsubsection{Ground Glass Diffusers}
Ground-glass diffusers are the simplest of the four diffuser types discussed here. These diffusers are generally produced by sand-blasting glass using various grit sizes to create small randomized surface features. As the surface features are randomized, ground-glass diffusers offer little control over their diffusing characteristics, resulting in limited control on angular divergence. As such, these diffusers are only capable of producing Gaussian intensity profiles. Although commercially available at low cost, these diffusers have low optical transmission efficiencies, and diffuse light at large angles. Ground glass diffusers with opal coatings can be made to achieve close to Lambertian diffusion.

\subsubsection{Holographic Diffusers}
Holographic diffusers rely on the holographic recording of a speckle pattern in the diffuser substrate. This speckle pattern creates pseudo-random semi-periodic surface structures that can be controlled in a statistical sense, offering precise control over the angular distribution of the output light. These types of diffusers are available from many vendors, including Edmund Optics, and notably Luminit LLC which offers holographic Light Shaping Diffusers\textregistered~which can be made to have very high transmission efficiencies of over 92\%\footnote{See holographic Light Shaping Diffusers at the Luminit LLC website: \url{http://www.luminitco.com/}}. However, as the surface structures are only controlled in a statistical sense, this limits the angular diffusion patterns to be either circular or elliptical, and only offers Gaussian-like intensity profiles \citep{sales2004}. This is suboptimal for high-precision aperture photometry, as the Gaussian profile has broad and extensive wings, spreading out the signal outside the photometric aperture. Furthermore, Gaussian-shaped PSFs have significant slopes across the full PSF except in the very center, making them more subject to guiding errors and changes in seeing.

\subsubsection{Diffractive Diffusers}
Diffractive diffusers are based on fabricating a phase mask for a single central wavelength, and can be made to have efficiencies between 80\% up to 90-95\%\footnote{See e.g.,~discussion on diffractive diffusers at the RPC Photonics website \url{http://www.rpcphotonics.com/product/diffractive-optics/}}. However, the output pattern is highly sensitive to the wavelength of light used, and as such, these diffusers are largely limited to monochromatic applications in laser systems, but could, however, have possible applications in narrow-band astronomical studies. Due to their high sensitivity with wavelength, we did not study these types of diffusers further for our broad-band photometry applications.

\subsubsection{Engineered Diffusers\texttrademark}
Unlike the other types of diffusers which only offer statistical control of the surface features, diffusers which precisely control the shape, size, and location of its surface features in a deterministic manner have the highest degree of control over their output. Such diffusers, capable of molding the output to a desired intensity profile and light distribution pattern, are now commercially available. We worked closely with RPC Photonics in Rochester New York, to test and design the diffusers used in this paper. These Engineered Diffusers\texttrademark offer precise beam control capabilities, and utility for many applications.

\begin{figure}[t]
	\begin{center}
		\includegraphics[width=\columnwidth]{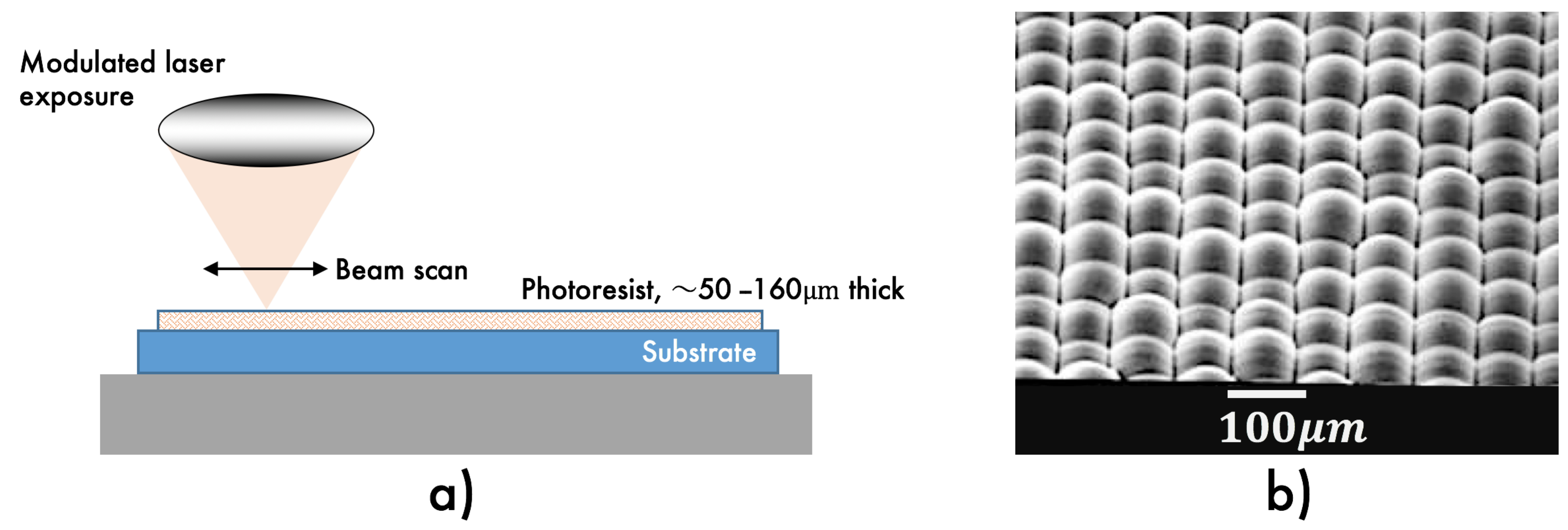}
	\end{center}
	\caption{a) A schematic diagram of the laser writing process: a modulated laser beam is scanned across a surface to deterministically write in surface features; b) Surface Electron Microscope image of the surface of an Engineered Diffuser\texttrademark, demonstrating a deterministic placement of surface features and microlenslets. Image used with permission from RPC Photonics.}
	\label{fig:laserwriter}
\end{figure}

Engineered Diffusers\texttrademark~\citep{rpcpatent} are composed of individually manipulated unit cells or microlenslets (Figure \ref{fig:laserwriter}). By precisely controlling the design and manufacturing process, a surface can be engineered to produce a desired intensity profile and light distribution pattern for a given input beam. To ensure that the diffuser output is stable towards varying beam input, the size, shape, and location of the microlenslets are varied according to a pre-defined probability distribution chosen to implement the desired beam shaping functions \citep{sales2004}. Additionally, this microlens distribution can be carefully designed to avoid discontinuities and minimize scattered light and diffraction artifacts from the output. In this manner, Engineered Diffusers\texttrademark~retain the best properties of both random and deterministic diffusers.

The diffusers we used are manufactured by RPC Photonics using a proprietary laser writing process (Figure \ref{fig:laserwriter}). The process starts by making a master, consisting of a substrate coated with a thick layer of photoresist. A focused UV laser beam is scanned across the surface, and by modulating the intensity of the laser beam, different exposures can be achieved, which breaks down the photoresist. By developing out the exposed areas, a deterministically structured surface with controlled size, shape and depth is produced (Figure \ref{fig:laserwriter}b). The master can then be used to produce sub-masters and replicas in different materials, including fused silica, silicon, or on polymer-layers on top of glass, allowing these diffusers to span a wide application wavelength range from 193 nm to 10.6 $\mu$m \citep{sales2004}.

Although the details of the exact design and manufacture of these structures are proprietary to RPC Photonics, a few general design rules are noteworthy:
\begin{itemize}
\item Designing and fabricating diffusers with larger angles of diffusion is easier than smaller angles, if a top-hat like shape with a steep fall off is desired. 
\item Polymer diffuser patterns bonded to glass substrates (like those we tested at the APO 3.5m) are less expensive since the laser writing process can be used to make and replicate them. Pure fused silica diffusers (like those we tested at Palomar) are made with a process similar to a reactive ion etch and are more time consuming and expensive.
\item Discontinuities in the surface structures can lead to additional scattered light. This can be mitigated in the design process if the requirements are well stated.
\item Systematic periodic errors or shifts in the laser writing process can result in the diffuser pattern to be grating-like, diffracting light at very low efficiencies into non-zero orders. This was discovered with tests on the ARC 3.5m (see Section \ref{sec:grating}). This can also be mitigated in future designs (RPC Photonics, private communication).
\end{itemize}

\subsection{Using diffusers in telescopes for precise photometry}

\begin{figure*}
	\begin{center}
		\includegraphics[width=0.8\textwidth]{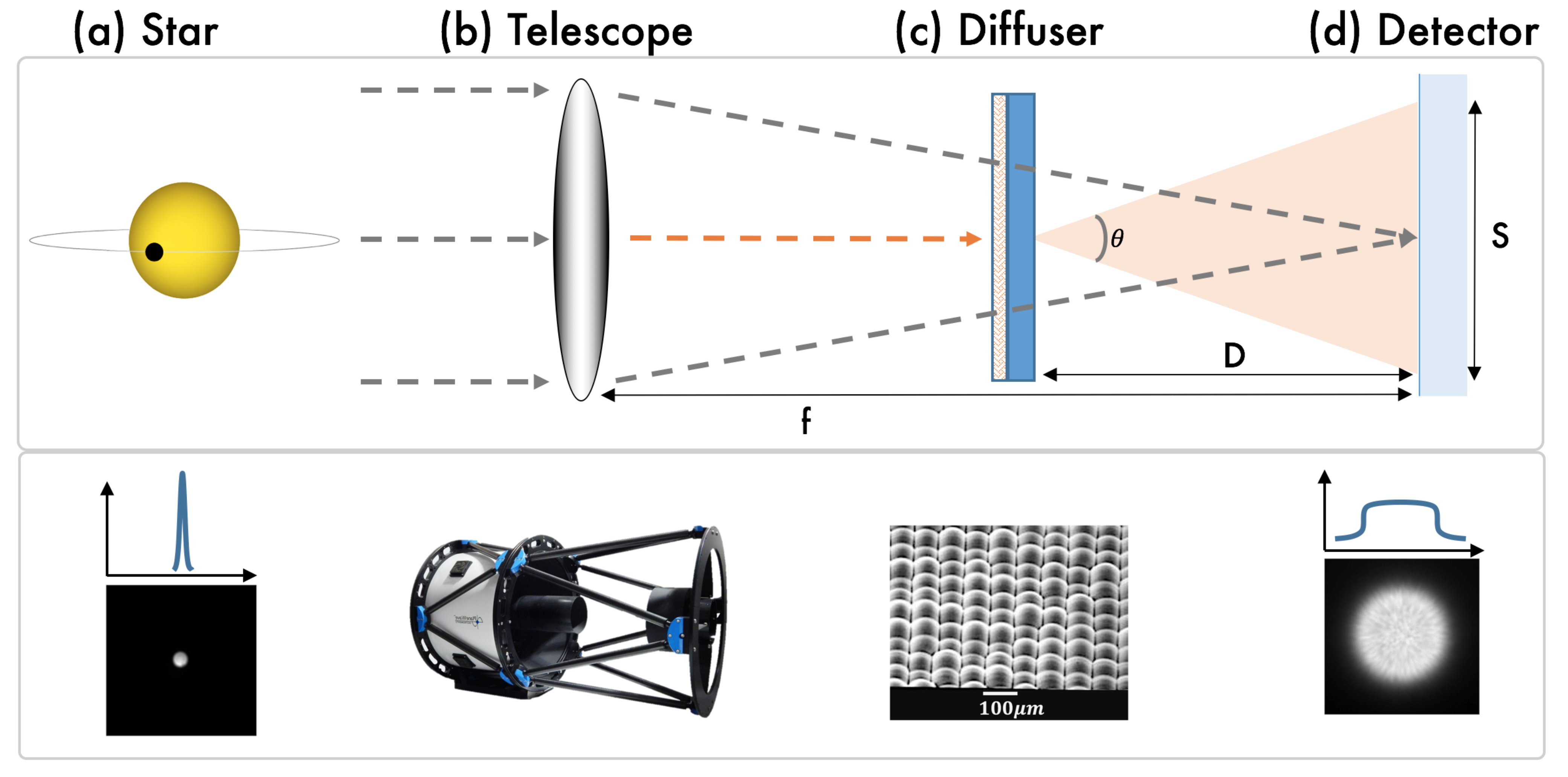}
	\end{center}
	\caption{Diffuser usage in a telescope in a converging beam. Microscopically engineered patterns on the surface of the diffuser (c) are used to mold starlight in a converging beam (a;b) to a broad and stable top-hat shape on the detector (d). Diffuser surface structures, image credit RPC Photonics.}
	\label{fig:diffuserintelescope}
\end{figure*}

Due to their light-shaping and beam homogenizing capabilities, beam-shaping diffusers are attractive optical devices to use for high-precision photometry applications. In the ideal case, such a diffuser would create a top-hat PSF shape with steep sides and a flat top subtending many tens of pixels in diameter, minimizing the signal lost outside the photometric aperture. Furthermore, like mentioned above, a top-hat PSF is more favorable than e.g., a Gaussian-shaped PSF, which has significant slopes everywhere except in the very center, making every pixel subject to guiding errors and changes in seeing. Meanwhile, a top-hat PSF restricts guiding errors due to PSF slopes to only the edge pixels, because the inner pixels see the same flux regardless. To enable the adoption of a beam-shaping diffuser over a broad range of astronomical instrumentation and allow for maximum flexibility, the diffuser should work in both converging and collimated beams. 

Figure \ref{fig:diffuserintelescope} shows a schematic of a diffuser in a converging telescope beam. Light from a star (a) arrives at the telescope (b) as a collimated beam. In this schematic, the telescope acts as a lens with an effective focal length $f$. The diffuser, with an opening angle of $\theta$ is placed in the converging beam at a distance $D$ from the detector image plane. The diffuser pattern faces the incoming starlight. The approximate FWHM of the resulting diffused spot on the image plane is given by:
\begin{equation}
S = D \tan{\theta}.
\label{eq:fwhm}
\end{equation}
By using different distances from the diffuser to the focal plane the size of the resulting PSF can be tuned. In most telescope systems, this translates to a relatively small opening angle $\theta$, on the order of 0$^{\circ}$.05-0$^{\circ}$.5. However, we have also tested angles as large as $2^{\circ}$ with excellent photometric performance.

\begin{figure}[t]
	\begin{center}
		\includegraphics[width=\columnwidth]{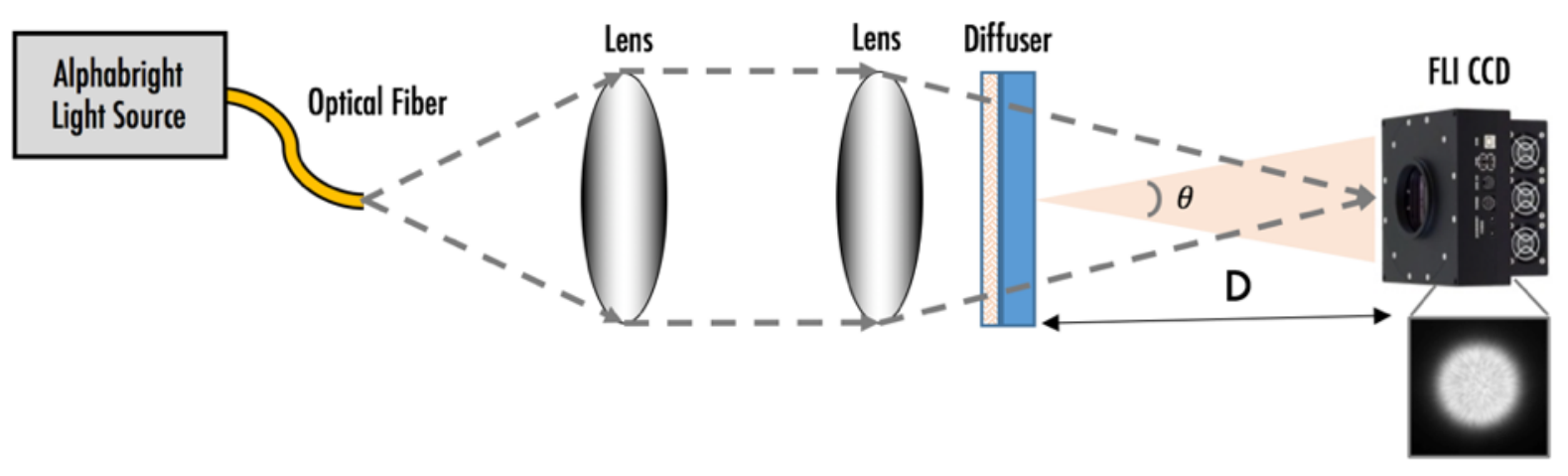}
	\end{center}
	\caption{Lab test setup to characterize diffusers in converging and collimated beams.}
	\label{fig:labsetup}
\end{figure}

\section{Lab Setup and Diffuser-Assisted Observations}

\subsection{Diffuser Characterization Lab Setup}
To study diffuser PSFs in collimated and converging beams, and how their PSF changes with distance, we set up a dedicated test bench (Figure \ref{fig:labsetup}). A single mode fiber was coupled to a collimating lens system, composed of two identical 2in diameter f/6.3 camera lenses. While coupling of a single mode fiber to a broad-band light source is inefficient, this posed no issues for our experiment. To better match the properties of incoherent starlight on-sky, we explicitly avoided using a coherent supercontinuum source in our lab setup in favor of a low-coherence broad-band source (Alphabright Quartz-Tungsten-Halogen light source). The diffusers were mounted in a custom-made rotation mount at a varying distance D from an FLI Proline PL4710-1MB monochrome 1056x1027 CCD camera with 13x13 micron pixels. Before mounting the diffuser, the test bench was aligned to ensure the optical system was properly focused. The rotation mount used a stepper motor driven by an Arduino, capable of rotating the diffuser at 1-2 Hz. This enabled us to smooth out the PSF for exposure times of a few seconds, as is discussed further in Section \ref{sec:opticalresults}. The exposures were dark subtracted, using a master median dark built from a sequence of 25 dark frames.

\subsection{Diffusers used}
Table \ref{tab:diffusers} summarizes the different diffusers tested in this paper. Most of the diffusers were Engineered Diffusers\texttrademark~fabricated by RPC Photonics. Additionally, we tested a holographic diffuser from Edmund Optics with a $1^{\circ}$ opening angle. However, being a holographic diffuser, it had a Gaussian intensity profile with broad extensive wings, suboptimal for precise aperture photometry. Instead, we focused further tests on the three top-hat off-the-shelf diffusers from RPC Photonics with opening angles $0^{\circ}.25$, $0^{\circ}.5$ and $2^{\circ}.0$, respectively. The off-the-shelf diffusers from RPC were not optimized for our application. Therefore, we worked with RPC to fabricate a custom diffuser optimized for installation in ARCTIC, with specifications informed from the lab tests of off-the-shelf diffusers. Specifically, we optimized the customized diffuser to give a top-hat PSF with a 10" FWHM on ARCTIC detector array, resulting in a diffuser angle of $0^{\circ}.4$. The final diffuser opening angle as fabricated was $\sim$0$^{\circ}$.34.

\begin{table*}
	\centering
	\caption{A summary of the different diffusers tested and used in this work.}
	\begin{tabular}{l l l l l}
	\hline\hline
	  Opening Angle & Company & Diffuser Part Number & Type & Size \\ \hline
      $1.0^{\circ}$   & Edmund Optics  & \#47-990              & Holographic Diffuser                       & 2"x2" \\
      $0.25^{\circ}$  & RPC Photonics  & EDC-0.25-07118-A-2S   & Off-the-shelf Engineered Top-Hat Diffuser\texttrademark  & 2"x2" \\
      $0.5^{\circ}$   & RPC Photonics  & EDC-0.5-07101-A-2S    & Off-the-shelf Engineered Top-Hat Diffuser\texttrademark  & 2"x2" \\
      $2.0^{\circ}$   & RPC Photonics  & EDC-2-07331-A-2S      & Off-the-shelf Engineered Top-Hat Diffuser\texttrademark  & 2"x2" \\
      $0.34^{\circ}$  & RPC Photonics  & Custom                & Custom Engineered Top-Hat Diffuser\texttrademark         & 150mm circle \\
      $0.08^{\circ}$  & RPC Photonics  & Custom                & Custom Engineered Top-Hat Diffuser\texttrademark         & 60mm circle \\
	\hline
	\end{tabular}
	\label{tab:diffusers}
\end{table*}

Our NIR diffuser was designed for the Wide-field Infra-Red Camera (WIRC) at the Palomar 200-in Hale telescope. WIRC sits at the prime focus of the telescope and consists of a collimation assembly that collimates the beam coming from the primary, followed by a Lyot stop and two filter wheels tilted by $7^{\circ}$ to minimize ghost reflections. Because the filter wheel assembly in WIRC is located in the collimated beam, we chose to install the diffuser directly into one of the filter slots inside the WIRC cryogenic dewar for minimal modification to the instrument. 

Our NIR diffuser was designed to satisfy four conditions: 1) the FWHM of the diffused PSF should be large enough to spread starlight onto a large number of pixels for the purpose of mitigating inter- and intra-pixel variations; 2) FWHM of the diffused PSF needs to be larger than that of the seeing PSF to ensure that the diffused PSF does not change significantly under variable seeing conditions; 3) the FWHM also needs to be small enough such that stars of $\sim$11-12 mag\footnote{Based on typical brightness of faint reference stars in our target field} in $K_S$ are not limited by background noise; 4) the FWHM needs to be small enough to avoid cross-talk in PSFs for neighboring stars. Given that the typical seeing at the Hale telescope is between 0.5"-1.5", we thus designed the diffuser to have a top-hat PSF with FHWM of 3" to meet the above conditions. Because the filter wheels are inside the cryogenic dewar, the diffuser is made of fused silica for its low thermal expansion coefficient. The size is 60mm x 5mm, compatible with the filter wheel in the camera. The working wavelengths of the diffuser were optimized to cover the NIR $J$, $H$, and $K$ bands.

\subsection{On-sky Diffuser-Assisted Observations}
Table \ref{tab:telescopes} describes the 3 observatories used to perform diffuser-assisted photometry in this work. The observations for each are further discussed below.

\begin{table*}
	\centering
	\caption{A table summarizing the different telescopes where we have tested diffusers on sky.}
	\begin{tabular}{l l l l l}
	\hline\hline
	           & Parameter                & CDK 24                                                      & APO            &  Palomar              \\ \hline
	  General  & Telescope                & PlaneWave CDK 24                                            & 3.5m ARC       &  Hale 200"            \\
	           & Instrument               & Apogee Aspen CG42                                           & ARCTIC         &  WIRC                 \\
	           & Beam f/\# at diffuser    & f/6.3                                                       & f/8.0          &  Collimated           \\
	           & Detector                 & CCD                                                         & CCD            &  Hawaii-2             \\
	           & Wavelengths              & Optical                                                     & Optical        &  NIR $J$,$H$,$K$      \\ \hline
	  Diffuser & Opening angle ($\theta$) & $0.25^{\circ}$, $0.5^{\circ}$, $2.0^{\circ}$ & $0.34^{\circ}$ &  $0.08^{\circ}$       \\
	           & Distance from detector   & 50mm                                                        & 200mm          &  (In collimated beam) \\
               & Diffuser size            & 2"x2"                                                       & 150mm circle   & 60mm circle           \\
			   & PSF FWHM                 & 12", 23", 92"                                               & $\sim$9"       &  $\sim$3"             \\
	           & Rotation capability      & No                                                          & Yes            &  No                   \\

	\hline
	\end{tabular}
	\label{tab:telescopes}
\end{table*}

\subsubsection{Penn State PlaneWave CDK 24"}
To verify the operation of diffusers, we tested various diffusers on-sky using the Penn State PlaneWave 24" telescope. The telescope was installed in 2014 at Davey Lab Observatory in University Park, Pennsylvania, at an altitude of 360m above see level. The telescope has an Apogee/Andor Aspen CG 42 camera, using a CCD42-10 2048x2048 pixel chip from E2V with 13.5micron pixels. This results in a plate scale of $\sim$0.77"/pixel, and a FOV of 24'x24'. The telescope is equipped with a dual filter wheel (AFW50-10S dual filter wheel), capable of housing 20 2"x2" filters. The diffuser was placed in the filter wheel closer to the camera, so the diffuser was 50mm away from the detector. Using a dual filter wheel allowed us to easily perform diffuser-assisted observations in different filters.

\paragraph{55 Cnc}
As an illustration of our observations with Penn State's CDK 24", we discuss our out-of-transit observations of 55 Cnc using a $2.0^{\circ}$ off-the-shelf diffuser on this telescope. 55 Cnc is a nearby bright (V=5.95mag) G8V binary star (its companion 55 Cnc B is an M4.5V dwarf). We used 53 Cnc, a nearby bright M3III star (V=6.23) star, $\sim$4.5' away from our target, as our main reference star. We observed the system in Johnson I, as both stars are well matched in brightness in that filter, and to minimize the impact from the Moon brightness, which was at $\sim$88\%. The observations were done on March 27th, 2016, from 04:00UT to 06:30UT. The target was setting, and starting at airmass 1.16, ending at airmass 1.9. The conditions were good, with little-to-no clouds. The $2.0^{\circ}$ diffuser allowed us to spread the PSF over $\sim$$120"$. Without the diffuser, the detector saturated almost instantly.

The exposure time was 120s, with a 11s dead-time between exposures. This allowed us to collect >$10^8$ counts in the target and reference star. We took 20 dome flats, and 20 dark frames, and median combined them using AstroImageJ \citep{collins2017} to create master dark and flat frame images. We used AstroImageJ for the photometric reduction. The aperture setting that gave the smallest residuals was 100, 150, 200 pixels, for the aperture radius, and the inner and outer annuli, respectively. To arrive at the final light curve, we detrended the raw data with airmass, a straight line, and $x$ and $y$ pixel centroid coordinates using the detrend function in AstroImageJ.

\subsubsection{Apache Point 3.5m telescope}
In September 2016 we installed a custom Engineered Diffuser\texttrademark~as a part of the Astrophysical Research Consortium Telescope Imaging Camera (ARCTIC) on the ARC 3.5m Telescope at Apache Point Observatory. ARCTIC uses a back-illuminated STA4150LN BI 4096x4096 pixel CCD with 15 micron pixels. This gives an unbinned plate scale of 0".114/pixel, and a field of view of 7'.5x7'.5. The detector has four amplifiers, and can be read out using one amplifier (lower left), or using all amplifiers simultaneously dividing the frame into 4 quadrants.

\begin{figure*}[t]
	\begin{center}
		\includegraphics[width=0.8\textwidth]{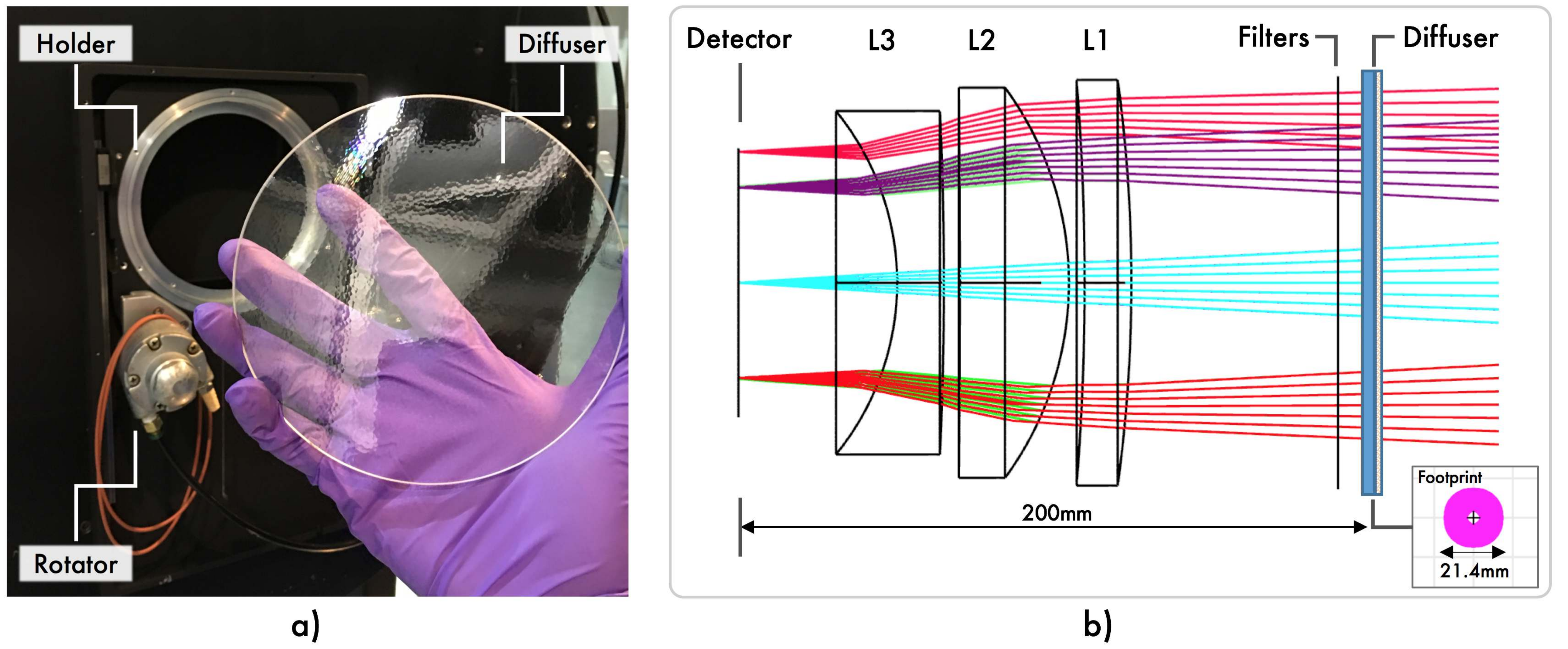}
	\end{center}
	\caption{a) An image of the diffuser at APO. The diffuser pattern is clearly seen. The diffuser is mounted on a retaining ring that can rotate on demand during exposures. B) A schematic diagram of the diffuser and its location in the ARCTIC imager. The rays shown are the ray-traces as calculated using Zemax OpticStudio without the diffuser. The footprint diagram shows the beam footprint at the diffuser location.}
	\label{fig:diffuseratAPO}
\end{figure*}

Figure \ref{fig:diffuseratAPO}a shows an image of the final diffuser at ARCTIC, along with the dedicated holder and rotator designed by us. The holder is capable of sliding the diffuser in and out of the telescope beam, and rotating the diffuser during observations with a pneumatic motor that moves with the diffuser holder assembly. The holder places the diffuser in front of the ARCTIC 6 position filter wheel, and $200 \unit{mm}$ away from the detector plane, creating a top-hat PSF with $\sim$9" FWHM. Figure \ref{fig:diffuseratAPO}b shows a schematic diagram of the diffuser location in ARCTIC, where the rays are traced with no diffuser in the beam path. The inset shows a footprint diagram of the beam at the diffuser location: an annulus due to the central obstruction of the telescope. These parameters---the size of the beam and the distance from the detector plane---were key parameters in the optimization process of the diffuser.

For some of our on-sky observations with ARCTIC and the diffuser we used an off-the-shelf narrow-band filter from Semrock operating in a band with minimal water absorption centered at 857nm (Figure \ref{fig:semrock}). However, the filter was only 2"x2" in size, truncating the FOV from 8'x8' to about 180"x180". Therefore, our on-sky tests of the diffuser in this filter were limited to bright, closely separated targets. We did not observe any noticeable fringing effects in this filter with ARCTIC, which is commonly seen with redder filters such as SDSS $z^\prime$.

\paragraph{16 Cygni}
Using ARCTIC with a diffuser, we observed 16 Cygni, a nearby bright G-dwarf binary star, where 16 Cyg A and B have V magnitudes of 5.95, and 6.2, respectively, and a separation of 39.5". The observations were performed on September 20th, 2016 from 2am to 6am UT. The target rose during the observations, starting at airmass 1.07, peaking at 1.05 at the meridian, and ended at 1.30 at the end of the observations. Moon illumination was $\sim$85\%. The night was not photometric, with variable seeing >2" FWHM and intermittent clouds. Nevertheless, due to the flux homogenizing properties of the diffuser, the PSF remained stable throughout the night. To maximize observing efficiency and minimize scintillation noise, the detector was configured to use one amplifier in fast readout mode, reading out only a subframe of 783x813 pixels encompassing 16 Cyg A and the reference star 16 Cyg B. This resulted in a short readout time of 5s.  Binning was set to 1x1 due to the brightness of the target, resulting in an exposure time of 16s to reach $\sim$40,000 peak counts per pixel for 16 Cyg A, and thus a total observing cadence of 21s, and duty cycle of 76\%.

Our data reduction consisted of standard aperture photometry performed with the AstroImageJ software suite. After experimenting with different aperture radii and annuli, the best RMS precision was achieved with an aperture radius of 100 pixels, and inner and outer annulus radii of 100 and 200 pixels, respectively. For calibration, a set of 25 bias frames were used, along with 25 dome flats. Each set was median combined using AstroImageJ. 

16 Cygni A and B were observed by \textit{Kepler} in both short and long-cadence mode, and in this paper, we compare our precision achieved with ARCTIC to that of \textit{Kepler}. There have been previous studies in the literature using \textit{Kepler} data of 16 Cygni A and B: using short cadence data for astroseismology \citep[e.g.,][]{metcalfe2012,lund2014} and gyrochronology \citep{davies2015}, and long cadence data to study the link between radial velocity variations and photometric flicker \citep{bastien2014}. 16 Cygni A, and B have Kp magnitudes of 5.864, and 6.095, and Kepler Input Catalog (KIC) IDs of KIC12069424 and KIC12069449, respectively. Due to their brightness, both stars are well above the saturation limit of \textit{Kepler}, which is $K_p\sim$11.5 \citep{gilliland2011}. Still, excellent precision levels can be achieved for saturated stars in the \textit{Kepler} data, due to the conservation of charge in the \textit{Kepler} CCDs, by summing up the counts in the surrounding pixels, commonly yielding precisions down to 40ppm for stars $\sim$ 7mag \citep{gilliland2010}. The data is easily retrievable from MAST\footnote{MAST: \url{http://archive.stsci.edu/kepler/data_search/search.php}}. Short cadence data is available for both stars for Quarters 6-15, with the exception of Quarter 13. Quarter 6 has a known problem with its photometric precision as it did not use an optimized aperture in the photometric retrieval, resulting in data of rather poor quality \citep{lund2014}. Therefore, we focused our comparison to \textit{Kepler}, by looking at both the short cadence, and long cadence data from \textit{Kepler} from Quarters 7-15, excluding Quarter 13.

\paragraph{Transit of WASP 85 A b}
WASP 85 A b was discovered by \cite{brown2014}, and observed by \textit{K2} in Campaign 1 (EPIC 201862715). The star is a G5 dwarf, with a V magnitude of 11.2, and Kp magnitude of 10.247. WASP 85 A forms a close visual binary (angular separation of 1.5") with a cooler and dimmer (V=11.9) K0 dwarf companion, WASP 85 B. With the diffuser PSF FWHM being $\sim$9", this close proximity of the two stars causes the stellar PSFs to completely overlap in the diffuser images. Despite the overlapping PSFs, we are able to recover very high photometric precisions, as we discuss in Section \ref{sec:opticalresults}.

Our transit observations of WASP 85 A b were performed on 31th of January from 08:30 to 13:00 UT. The exposure time was initially set at 7s. However, the target was rising, starting from airmass 1.20 during the beginning of the observations, peaking at 1.11, and ending at airmass 1.54, and the exposure times were reduced to 6s after 15 minutes of observations to keep the exposures at $\sim$30,000 peak counts, well within the linear regime of the detector. This change in exposure time did not result in a visible change in photometric precision. Moon brightness was $\sim$12\%. The filter used was SDSS $r^\prime$. The binning mode was 4x4, with the detector read out in quad amplifier and fast readout mode, resulting in a readout time of 2.5s. We assume a total cadence of 8.5s, as $\sim$95\% of the images were taken at this cadence, resulting in an observing efficiency of 80\% for these observations.

The data reduction was performed using two independent photometry pipelines. First, we used a photometry pipeline being developed B. Morris et al.~(in prep), which implements principal component analysis (PCA) to find an optimal set of aperture radii, comparison stars, and environmental measurements used for detrending. The aperture radius used was 19 pixels, and the radius of the inner and outer radii were 32 and 55 pixels, respectively. The data were independently reduced using AstroImageJ using the same radii and detrending parameters, giving consistent results. A set of three reference stars within the FOV with median fluxes between 0.2-1.0 times that of the target were used in the differential photometry. A set of 22 darks, and a set of 20 flats were median combined to create a master flat and dark frames. In 4x4 binning at ARCTIC, a relatively high fraction of cosmic rays and charge particle events is observed in the science frames. To reduce the effect of these events on our photometry, we ran the data through the \texttt{astroscrappy} package\footnote{\texttt{astroscrappy} is available on Github here: \url{https://github.com/astropy/astroscrappy}}, a cosmic-ray rejection package written in Python, based on the Laplacian-edge cosmic ray rejection algorithm described by \cite{vanDokkum2001}. Using the default parameters in \texttt{astroscrappy}, resulted in fewer saturated outliers observed in the light curve.

To compare our best fit planet parameters with the values reported in the literature, we fit the transit using a Marcov Chain Monte Carlo (MCMC) approach described in Section \ref{sec:mcmc}.

\paragraph{Transit of TRES 3 b}
TRES 3 b is a hot Jupiter ($R\sim 1.3R_{\mathrm{Jup}}$) discovered by \cite{odonovan2007}, and is in a $P=1.306$ day orbit around a G4 V dwarf star with a V magnitude of 12.4. This target has been well studied in the literature \citep[e.g.,][]{knutson2014,sozzetti2009}. Beneficial for high-precision ground-based differential photometry, the TRES 3 field has a number of similarly bright reference stars close by (within the ARCTIC FOV for our purposes), which is beneficial for high-precision ground-based differential photometry. Choosing to observe this target thus allowed us to further test the limits of the diffuser-assisted photometry technique with ARCTIC, by observing a clear transit signal on a clear night at good airmasses.

Our observations of this target were performed on March 12th 2017 from 08:45 UT and 12:20 UT, where the target rose during the night, from an airmass of 1.90 to 1.04. The Moon was full during these observations (brightness $\sim$100\%), and thus to minimize sky background noise, we observed the target in SDSS $i^\prime$. We used an exposure time of 30s with ARCTIC in quad amplifier fast-readout 4x4 binning mode, resulting in a readout time of 2.5s. The total cadence was thus 32.5s, yielding an observing efficiency of 92\%. 

Similar to the WASP-85 observations above, the data reduction was performed using AstroImageJ, after cleaning up the images from cosmic rays and charged particle events using \texttt{astroscrappy}. We used 13 reference stars with a flux between 4\% and 180\% of the flux of the target star. After systematically testing a number of different aperture settings in AstroImageJ, the aperture radius that gave the smallest unbinned RMS scatter was 19 pixels, and with a radius of the inner and outer radii of 32 and 50 pixels, respectively. To create the final light curve we fit the transit guided with the parameters presented in \cite{odonovan2007}, performing a simultaneous transit-fit and detrending using AstroImageJ, using airmass, a straight line, and $x$, $y$-centroid pixel coordinates in AstroImageJ.

Similar to the WASP 85 A b transit, to compare our best-fit TRES-3b planet parameters with the values reported in the literature, we fit our TRES-3b observations with a MCMC approach described in Section \ref{sec:mcmc}.

\subsection{Palomar Hale/WIRC}
The Palomar experiment started in 2013. RPC delivered the first engineered diffuser with a Gaussian profile instead of a top-hat\footnote{Another diffuser closer to top-hat shape was remade and delivered within a few months}. We therefore went ahead and installed the diffuser on the WIRC camera in 2013 November for testing observations. 

Our on-sky test took place on UT 2013 December 21, using the old science grade HAWAII-2 array in WIRC with a wide FOV of 8.7'x8.7'. Unfortunately, due to a fatal failure (explosive debonding and separation of the semiconductor from its substrate) of the HAWAII-2 array a few months after our first on-sky test, we were not able to conduct additional tests as the replacement array was not science-grade, which significantly limited the precision of our photometry due to excessive hot and bad pixels and uneven linearity in different quadrants. Therefore, in this paper, we only demonstrate the performance of our IR diffuser using one night of observation. Since then, in late 2016 and early 2017, WIRC has been retrofitted with a science-grade HAWAII-2 detector, enabling us to continue our efforts in using a new diffuser on-sky using the updated WIRC system.

\begin{figure*}[t]
	\begin{center}
		\includegraphics[width=0.8\textwidth]{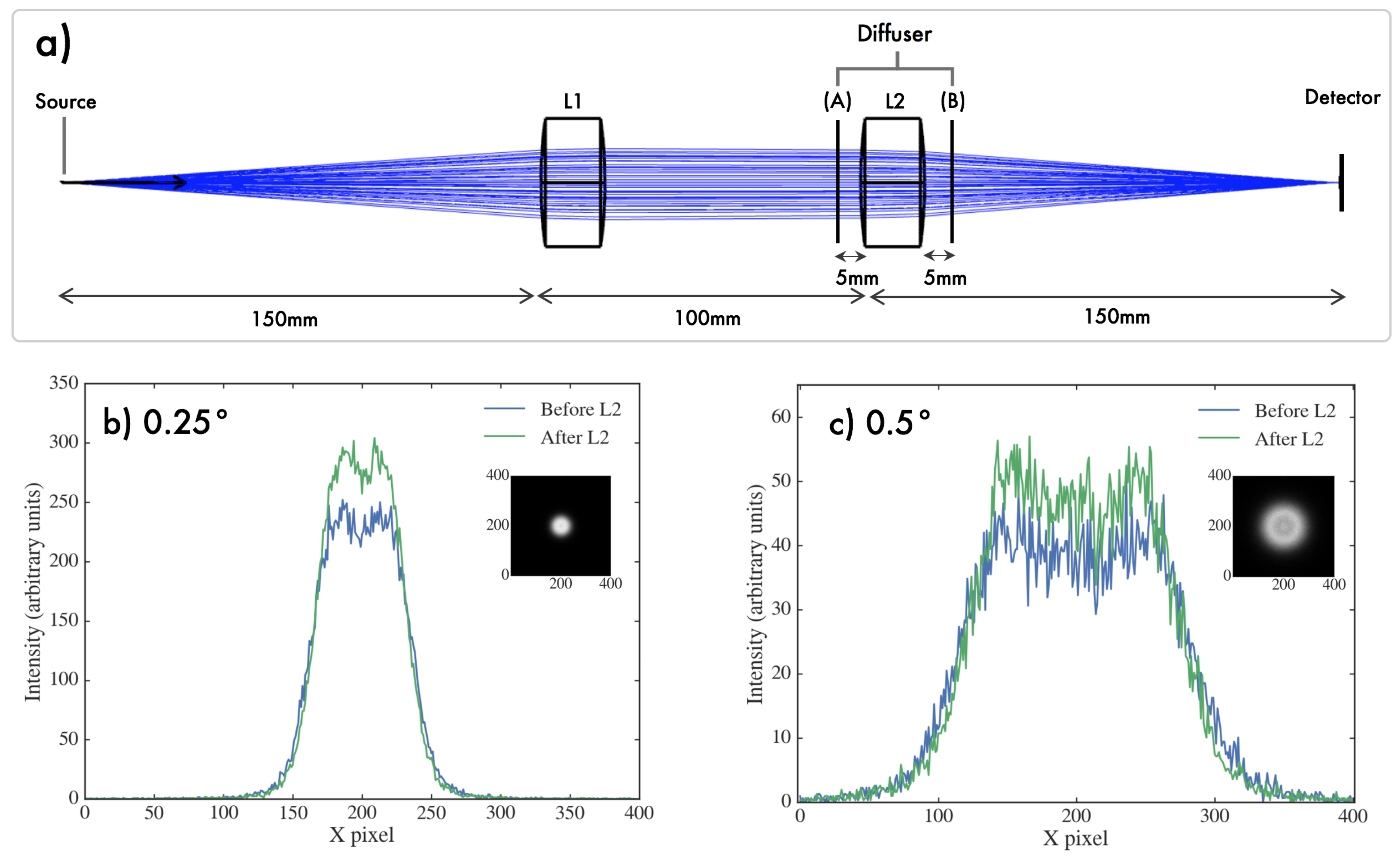}
	\end{center}
	\caption{Zemax simulations of diffusers: a) optical model used, consisting of a point source, two identical lenses (L1 and L2), and focused beam at a detector. B) Comparison of the resulting PSF when the diffuser is placed in the collimated beam (placement A; blue curve), and converging beam (placement B; green curve) for an off-the-shelf $0.25^{\circ}$ off-the-shelf diffuser. C) Same as B), except for a $0.5^{\circ}$ off-the-shelf diffuser.}
	\label{fig:zemax}
\end{figure*}

The observations were carried out in the $K_S$ band near the field of XO-3. Because XO-3 (K=8.8) is too bright to have enough reference stars of similar magnitude, we chose a nearby field and a fainter star, 2MASS J04230271+5740319 (K=10.79, V=13.26) as our photometry target. The observation lasted for only $\sim$3.5 hours before telescope closure due to critical weather conditions. Nonetheless, these observations still provided a useful test of the diffuser's performance. The telescope was kept focused during the observation while the diffuser was used to control the PSF. The final diffused PSF had a Gaussian shape with a FWHM of 17.4 pixels (4.35\arcsec) on average. A total number of 225 images were recorded continuously with 40s exposures, although 4 images were later rejected due to passing clouds, and using one double-correlated sampling (1 Fowler). 

Reduction of the images was carried out using our standard WIRC photometry pipeline described in \cite{zhao2012,zhao2014}. We corrected for time-varying telluric and instrumental effects by selecting 10 reference stars that had median fluxes between 0.3-1.0 times that of the target and showed no peculiarities in their light curves. Fainter stars were excluded due to low signal-to-noise, while stars brighter than the target saturated the detector. The $x$ and $y$ positions of the stars' centroids varied by less than 3 pixels, with a standard deviation of 0.63 pixel in $x$ and 0.51 pixel in $y$. The airmass changed from 1.22 at the start to 1.11 at the end. We applied 48 different aperture sizes with a step of 0.5 pixel for the target and reference stars, with the same sized aperture used for all stars in each step. The aperture with a radius of 16 pixels (4\arcsec) produced the smallest scatter in the reduced light curve, and was thus used for subsequent analyses.

\section{Results}
\subsection{Characterizing Diffusers}
\subsubsection{Diffusers work similarly in converging and collimated beams}
Because an imaging system inherently has a converging beam before the detector, the most straightforward way to incorporate a diffuser is in such a beam. However, some telescope systems have locations with collimated beams, where placing a diffuser would be more optimal. To compare the resulting PSF of diffuser placed in a converging versus a collimated beam, we modeled off-the-shelf diffusers using Zemax Opticstudio in non-sequential mode. For these simulations, we used Bidirectional Scattering Distribution Function (BSDF) data files available from the RPC Photonics website measured for off-the-shelf diffusers illuminated with on-axis input beams.

Figure \ref{fig:zemax}a shows an image of the optical model: an imaging system consisting of a point source emitting an F/6.3 beam at 550nm, a collimating lens (L1), and an identical lens (L2) to reimage the beam on a detector. Using this model, we studied the output at the detector by placing a diffuser in the collimated space before L2, and also in the converging beam after L2.

\begin{figure*}[t]
	\begin{center}
		\includegraphics[width=0.8\textwidth]{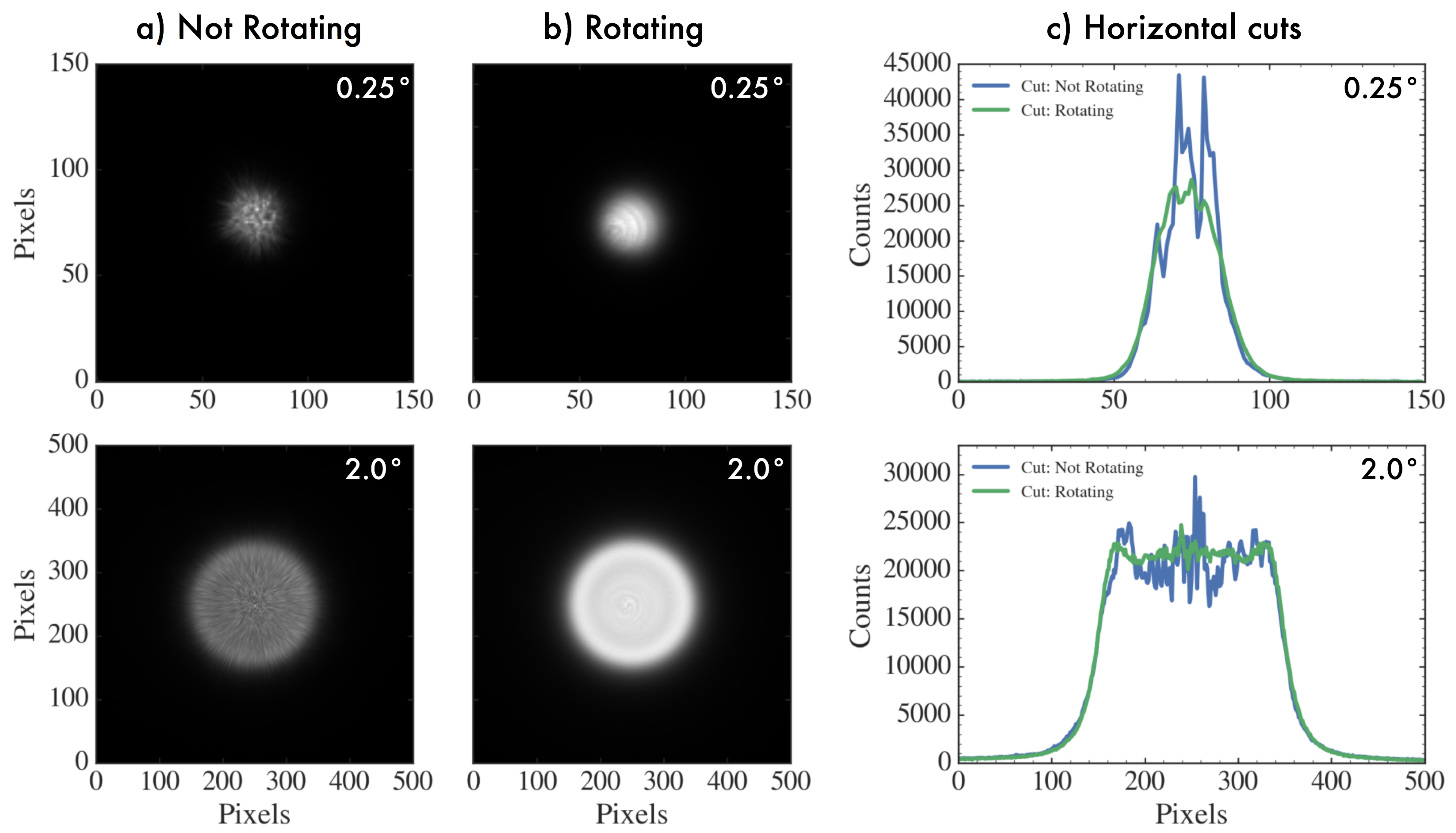}
	\end{center}
	\caption{A comparison of non-rotated (a) and rotated (b) diffused PSFs, for two off-the-shelf diffusers from RPC Photonics with an opening angle of $0.25^{\circ}$ and $2.0^{\circ}$, respectively. The panels in (c) show horizontal cuts through the center of the PSFs. The images in a) and b) have the same linear stretch.}
	\label{fig:rotationpanel1}
\end{figure*}

Figure \ref{fig:zemax} b) and c) show the resulting horizontal cuts on the detector for an off-the-shelf diffuser with opening angles of $0.25^{\circ}$, and $0.5^{\circ}$, respectively. Also shown in the insets are the respective modeled PSFs. For both diffuser opening angles, we see that the PSF shape, size, and speckle structures (high intensity peaks in the image plane) are similar for a diffuser placed in the collimated and converging beams. However, we observe that the intensity for the diffuser in the converging beam is about 20\% higher in both cases. This is due to two reasons. First, in the collimated beam the diffuser is slightly further away from the detector, broadening the resulting PSF. Second, with the diffuser in collimated space, the diffuser causes the incident rays on the L2 lens to be slightly diverging rather than collimated, moving the original focus position further away. Although not specifically shown in Figure \ref{fig:zemax}, this can be minimized by optimizing the focus of L2 after placing the diffuser in the collimated beam. With the diffuser in the collimated beam, an additional precaution will be to ensure that the now-diverging beam does not get clipped at the lens, i.e. that L2 has a large enough clear aperture to accommodate the larger beam footprint.

\subsubsection{Diffusers can be rotated to smooth out speckling}
The diffuser can be rotated during observation to smooth out the speckle pattern observed in diffused PSFs (Figure \ref{fig:rotationpanel1}). Being a statistically varied pattern of engineered structures, illuminating different parts of the diffuser will result in slight variations in the resulting output PSF. This PSF variation can be averaged over time by moving the diffuser during an exposure. Therefore, by taking an exposure that is longer than the characteristic time of change in the residual PSF variation, the resulting PSF can be effectively smoothed out.

In practice, the least design intensive path to move the diffuser in a beam is to rotate it along an axis parallel to the optical axis. We tested this in the lab: continuously rotating the diffuser at $\sim$1-2Hz effectively smoothed the output PSF, by allowing the diffuser to complete a few full rotations for exposure times of a few seconds. Figure \ref{fig:rotationpanel1} compares our lab measurements of rotated and non-rotated PSFs for two top-hat off-the-shelf diffusers from RPC Photonics with a $0^{\circ}.25$ and $2^{\circ}.0$ opening angle, respectively. Comparing the two non-rotating diffuser PSFs, we see that the amplitude of the speckles is larger for the $0^{\circ}.25$ diffuser than for the $2^{\circ}.0$ diffuser, or about 40\%, and 5-10\% of the total intensity, respectively. This results from the fact that it is easier to suppress speckling for larger opening angles.

Although not as good as allowing the diffuser to complete a few revolutions per exposure, we note that excellent PSF smoothing can already be achieved with only $\sim$$180^{\circ}$ of rotation during an exposure. Therefore, considering that our shortest exposure times for ARCTIC for high-precision photometry on bright stars are on the order 1-2s---with most exposure times for high-precision photometry being longer than 10s to maintain observing efficiency---the ARCTIC diffuser rotator was designed to produce a smooth rotational speed of at least 2-3Hz.

Small residual ripples with amplitudes of $\sim$2-10\% of the total intensity are seen in the rotated PSFs in Figure \ref{fig:rotationpanel1}. These ripples are concentric around the diffuser rotational axis. We see that for the $2^{\circ}.0$ diffuser exposure (Figure \ref{fig:rotationpanel1}) the rotational axis was more closely centered than in the $0^{\circ}.25$ diffuser exposure. However, as shown in the horizontal cuts in Figure \ref{fig:rotationpanel1}c these ripples are small in comparison to the spikes before rotating.

We note that the diffusers in Figure \ref{fig:rotationpanel1} are marketed as general top-hat diffusers with a specific opening angle. As such, they are advertised to provide a top-hat PSF over a broad parameter space, and were not specifically optimized for this test setup, i.e., to have a flat speckle pattern and steep wings. To find a diffuser more suited to our specific needs, we worked with RPC Photonics to design a custom top-hat diffuser that had better top-hat characteristics.

\begin{figure*}[t]
	\begin{center}
		\includegraphics[width=0.9\textwidth]{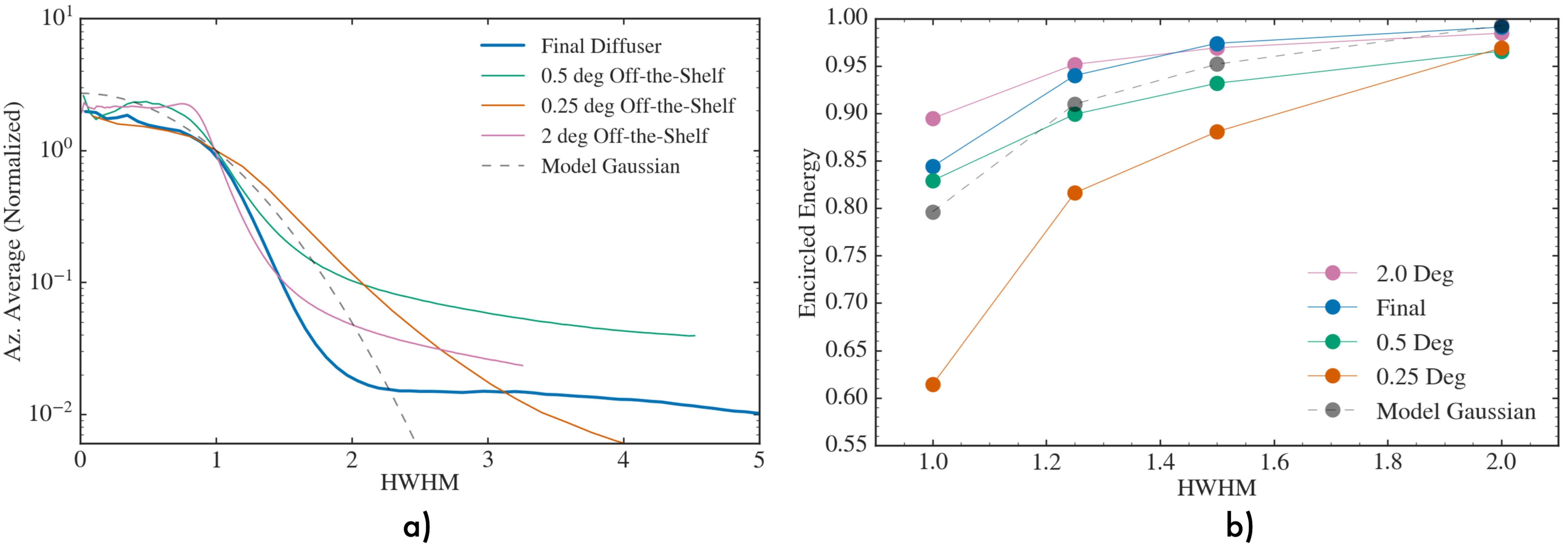}
	\end{center}
	\caption{A PSF comparison between off-the-shelf diffusers and our customized diffuser pattern. Also shown is a model perfect Gaussian PSF without noise. a) Comparison of azimuthally averaged PSFs, normalized to be 1 at the HWHM. b) A comparison of encircled energy as a function of HWHM, where EE(3HWHM) = 1. Our customized diffuser pattern is optimized to have a steep falloff and a flat top.}
	\label{fig:rotationpanel2}
\end{figure*}

\begin{figure*}[t]
	\begin{center}
		\includegraphics[width=0.9\textwidth]{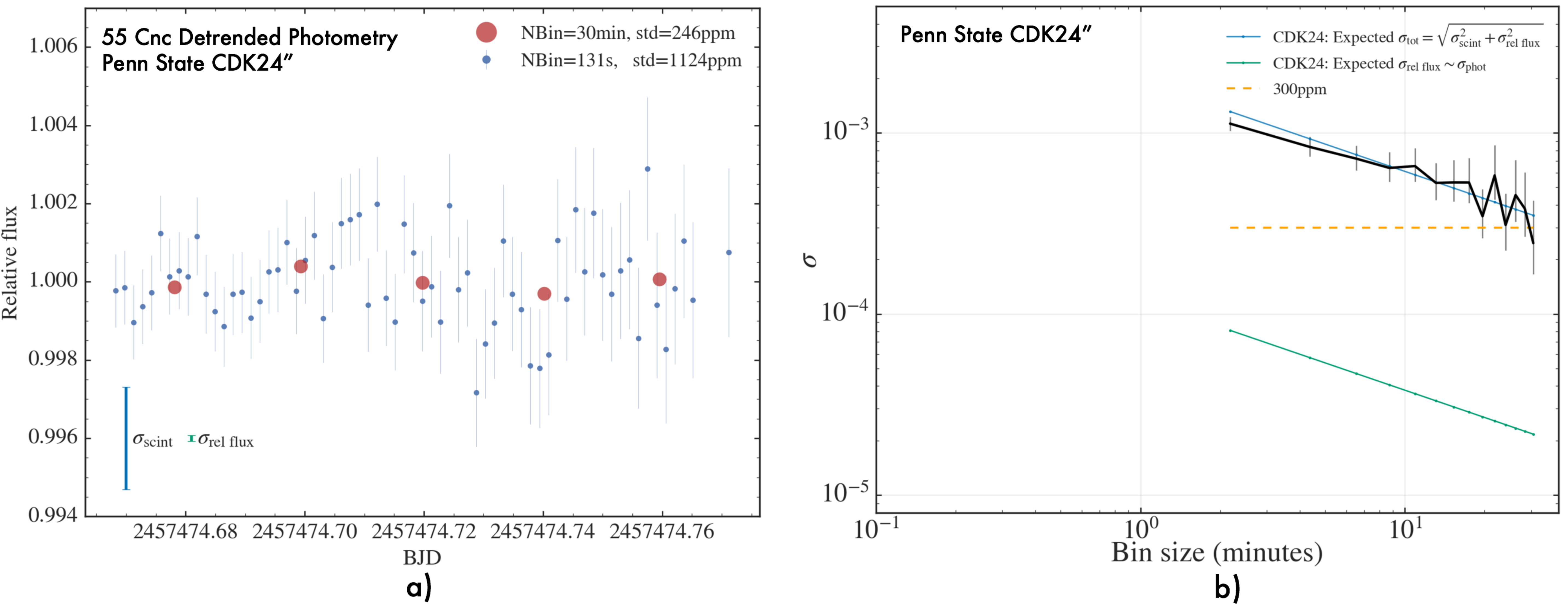}
	\end{center}
	\caption{a) A light curve of 55 Cnc (out of transit) as observed by the Penn State PlaneWave 24" in the Johnson I filter using a $2^{\circ}.0$ diffuser. Using the diffuser allowed us to observe this bright star with a cadence of 131s, resulting in an unbinned precision of 1124ppm. The photometric precision increases to $246_{-81}^{+176}$ppm in 30 minute bins. The errorbars are the total noise errorbars as calculated by Equation \ref{eq:stdtot}. Example scintillation and photon noise errors are shown in the bottom left corner in blue and green, respectively. b) Photometric precision as a function of bin size in minutes for our 55 Cnc observations (black). The green and blue curves show the expected calculated photon noise, and total photometric noise, as calculated using Equations \ref{eq:stdflux} and \ref{eq:stdtot}, respectively. We see that these observations are scintillation limited. A machine readable table including the photometry in panel a) is available in the online journal.}
	\label{fig:cdk24phot}
\end{figure*}

Figure \ref{fig:rotationpanel2}a compares the azimuthally averaged PSF of our custom developed diffuser pattern to those of commercial off-the-shelf diffusers. To facilitate the comparison, Figure \ref{fig:rotationpanel2}a shows azimuthally averaged PSFs normalized to be equal to unity at the HWHM of each PSF. We also show a model Gaussian with no background noise or background light. These data were taken at a fixed distance of $106 \unit{mm}$ away from the detector. From Figure \ref{fig:rotationpanel2}a, we see that our customized diffuser pattern provides PSFs with steeper wings than produced by off-the-shelf diffusers. We see that for our customized diffuser the PSF plateaus at a relatively constant level at $\sim$2HWHM, indicating that most of the signal for this diffuser will be within $\sim$2HWHM. The other diffusers fall off less steeply, and do not show evidence of a plateau in the range tested.

Figure \ref{fig:rotationpanel2}b shows the encircled energy (EE) (curve of growth) for the same set of diffusers tested. Due to the size of the detector used, we could only record up to $\sim$3HWHM of the PSF of the $2^{\circ}.0$ diffuser. Therefore, to compare the EEs, we set the EE(3HWHM) = 1 for all of the PSFs, and compare how quickly the EE or signal strength grows as a function of HWHM. We see that the customized diffuser pattern is better than the model Gaussian, and both off-the-shelf diffusers with a similar opening angle, both in terms of steepness of fall-off, and in terms of encircled energy. In this comparison, the $2^{\circ}.0$ diffuser is observed to have better encircled energy and a flatter top-hat, but is still to have some residual power towards the higher angles (falloff of wings could be steeper). As mentioned above, it is easier to fabricate and design diffuser patterns producing closer to ideal top-hats for diffusers with larger opening angles. However, for our astrophysical applications we needed diffusers closer to or smaller then $\sim$$0^{\circ}.4$, to achieve PSFs with a FWHM $\sim$10' with ARCTIC, to minimize sky background noise, and source overlap effects.

\subsection{Diffusers on-sky}
\subsubsection{Optical Diffuser-assisted photometry with Penn State's CDK 24}
Figure \ref{fig:cdk24phot} shows a result from some of our early tests with off-the-shelf diffusers on a small telescope. Specifically, Figure \ref{fig:cdk24phot} shows 2.5 hours out-of-transit observations of 55 Cnc using Penn State's CDK 24 telescope with a $2.0^{\circ}$ diffuser off-the-shelf diffuser from RPC Photonics. Using the diffuser allowed us to spread the PSF over a a large number of pixels, 160px FWHM, corresponding to $\sim$$20,000$ pixels illuminated per target. Spreading out the PSF over so many pixels, allowed us to increase our exposure time to 120s, and giving a final cadence of 131s (11s readout time), allowing us to achieve a duty cycle of 90\%, and gather $\sim$$10^8$ electrons in the target, and reference star apertures, respectively. The PSF remained stable throughout the observations. The resulting unbinned precision after detrending with airmass, $x$ and $y$ pixel positions was 1124ppm (see Figure \ref{fig:cdk24phot}).

Figure \ref{fig:cdk24phot} shows how the 55 Cnc observations bin down with increasing bin sizes (black curve). We calculate the error bars on our photometric precision, using the code described in \cite{cubillos2017}, assuming that for the highest binning sizes, the RMS scatter as a function of bin size follows an inverse-gamma distribution. The data bin down in a Gaussian-like manner, indicative of minimal residual systematic effects. In 30 minute bins, the data bin down shows a scatter of $246_{-81}^{+176}$ppm. This precision is similar to the precision needed to successfully observe the transit of 55 Cnc e \citep{mcarthur2004,dawson2010}---having a transit depth of 380ppm \citep{winn2011}---in a single night of observations. Although we note that observing only one transit at this precision would result in a marginal detection of the transit, the transit could be more precisely be characterized by co-adding a few transit observations to further increase the precision. This demonstrates that addition of such diffusers could significantly improve detection thresholds of transiting planets orbiting bright stars.

Furthermore, Figure \ref{fig:cdk24phot} compares the relative photometric flux errors, $\sigma_{\mathrm{rel\ flux}}$ (green curve), to the expected total photometric noise (including scintillation noise), $\sigma_{\mathrm{tot}} = \sqrt{\sigma_{\mathrm{rel\ flux}}^2 + \sigma_{\mathrm{scint}}^2}$ (blue curve), as calculated using Equations \ref{eq:stdflux}, and \ref{eq:stdtot}, respectively. For the airmass term in Equation \ref{eq:stdtot}, we assumed a fixed airmass equal to the mean airmass of the observations, amounting to $\chi$$\sim$1.4, which matches well with the observed scatter. From Figure \ref{fig:cdk24phot}, we see that the total noise is much larger than the photometric noise by an order of magnitude ($\sigma_{\mathrm{tot}} \gg \sigma_{\mathrm{rel\ flux}}$), where the total error is dominated by the scintillation noise. This is expected for a 24" (0.6m) telescope observing such a bright star from Figure \ref{fig:scint}.

\subsubsection{Optical Diffuser-assisted photometry with ARC 3.5m}
\label{sec:opticalresults}
\begin{figure}
	\begin{center}
		\includegraphics[width=\columnwidth]{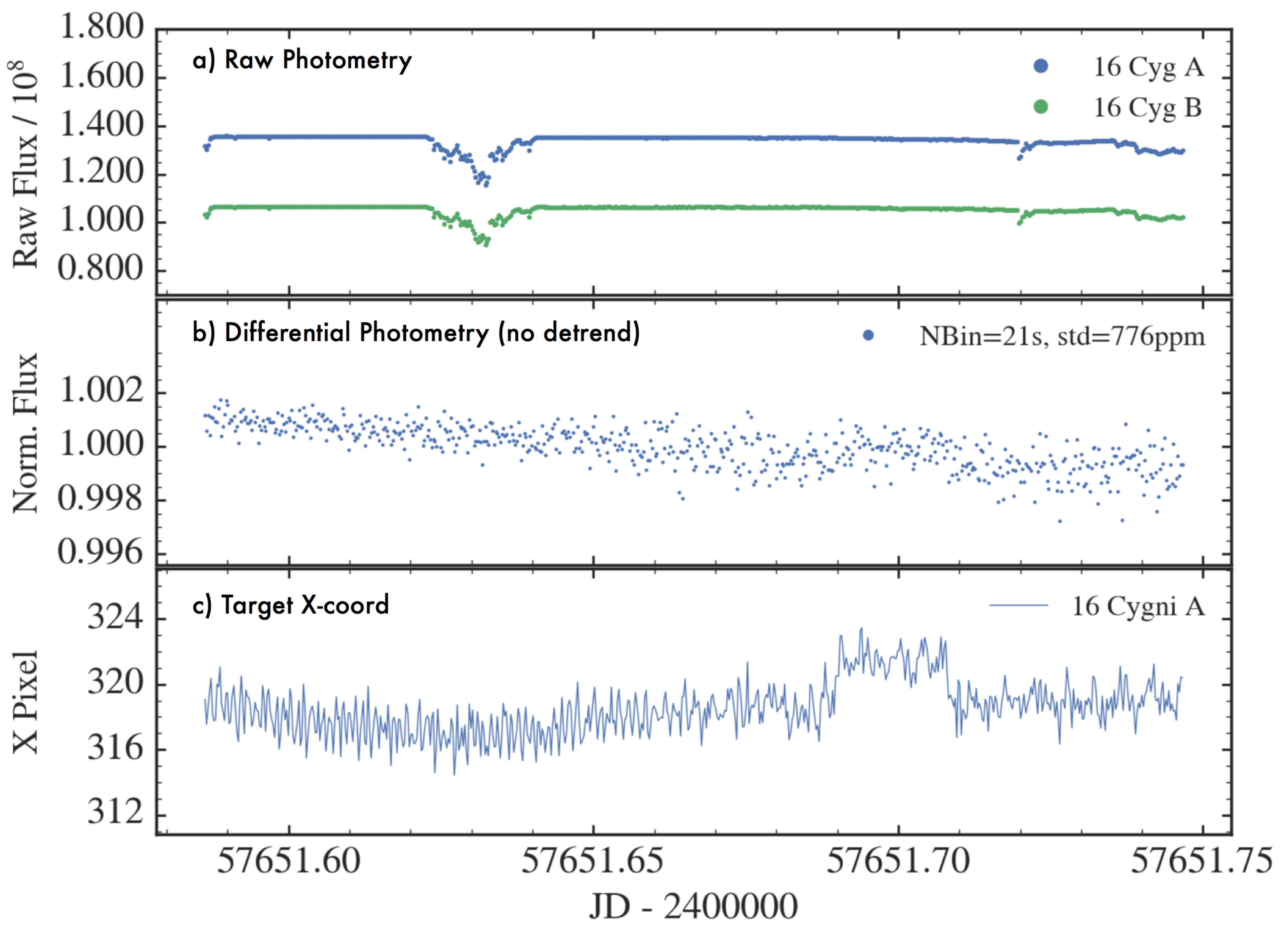} 
	\end{center}
	\caption{Demonstration of diffuser-assisted photometry in the optical: Engineering-time observations of the bright binary star 16 Cyg A. A) The raw photometry of 16 Cyg A, and 16 Cyg B. Clouds are seen after $\sim$1 hour from the start of the observations, and also towards the end. B) Normalized unbinned photometry of 16 Cyg A, without detrending. C) $x$-pixel coordinate of the target star 16 Cyg A as a function of time.}
\label{fig:cygniphotometrypanel}
\end{figure}

\begin{figure}
	\begin{center}
		\includegraphics[width=\columnwidth]{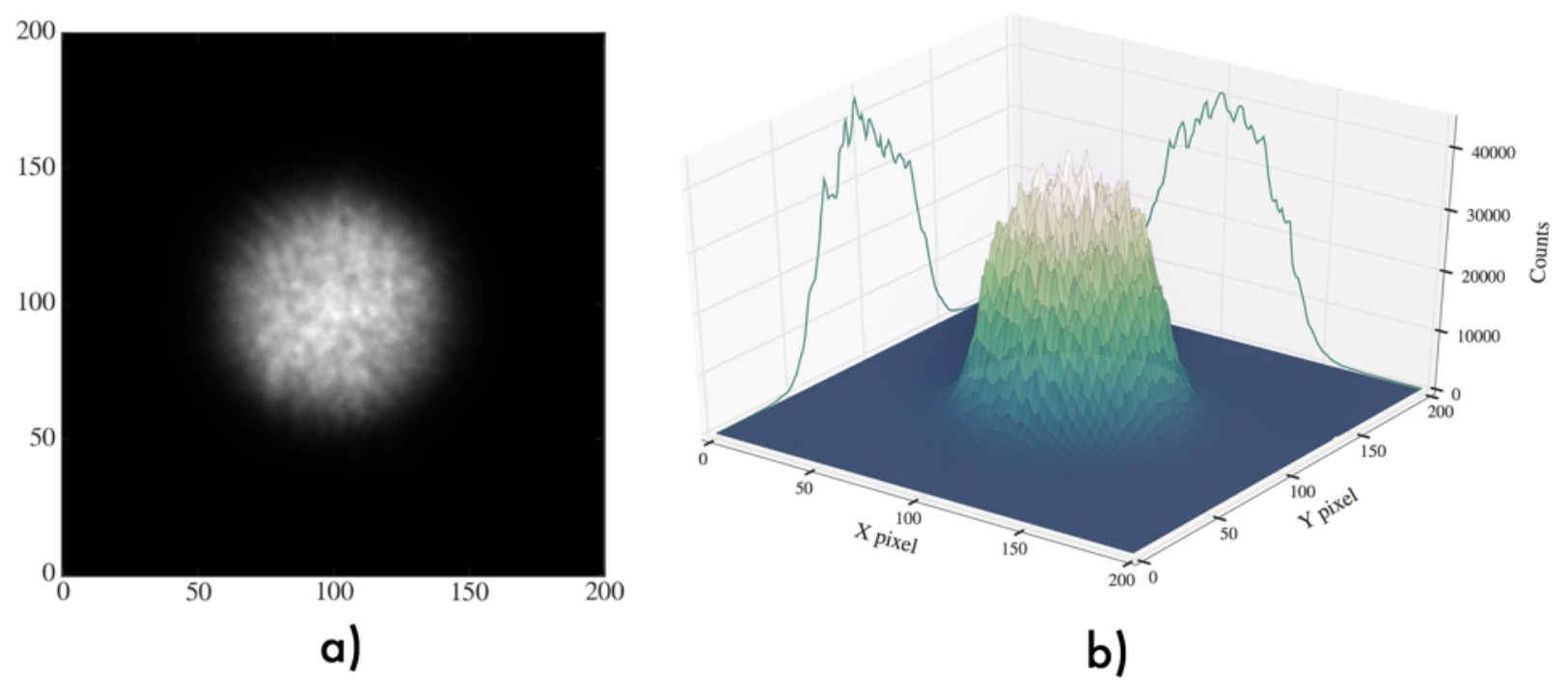}
	\end{center}
	\caption{3D view of a diffuser image chosen at random from the 16 Cyg A photometry. The ARCTIC diffuser rotator was not working reliably during this observing run, and this image shows the PSF when the rotator was not rotating, showing the small-amplitude speckle pattern, which remained stable throughout the observations. A video version of this figure can be found in the online version of the manuscript. The video demonstrates that overall the PSF remains stable throughout the observations. The smoothing of the residual speckle pattern seen in the video (e.g., towards the end of the video) is due to a combination seeing effects, and the rotator intermittently working.}
\label{fig:3dview}
\end{figure}

\begin{figure*}
	\begin{center}
		\includegraphics[width=0.9\textwidth]{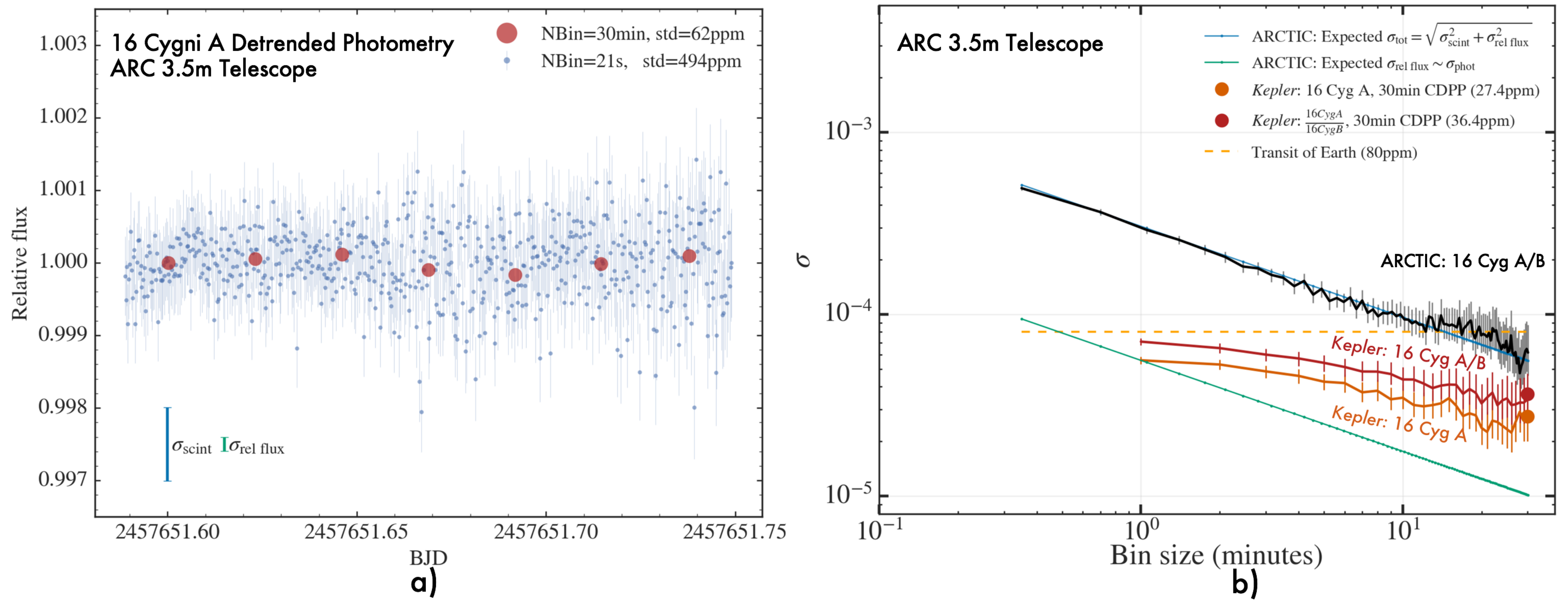}
	\end{center}
	\caption{a) Final detrended photometry of 16 Cyg A including scintillation errors, showing both unbinned (blue points), and in 30 min bins (red points); b) Comparing results with \textit{Kepler}: photometric precision as a function of bin size in minutes, comparing the 16 Cyg A diffuser data to a representative 4-hour short-cadence data of 16 Cyg A from \textit{Kepler} (orange curve; see details in text). Additionally, as a head-to-head comparison with our ground-based differential photometry, we show the \textit{Kepler} differential photometry for 16 Cygni A and B (red curve), which adds a $\sqrt{2}$ error in dividing the two lightcurves together. The bin size on the x-axis accounts for the full effective observing cadence (including both the exposure time and the dead time). The diffuser photometry reaches $62^{+26}_{-16}$ppm precision in 30 minute bins, a factor of $\sim$2 from \textit{Kepler}. A machine readable table including the photometry in panel a) is available in the online journal.}
\label{fig:16cygnipanel}
\end{figure*}

\paragraph{Photometric observations of 16 Cygni}
To demonstrate the photometric precision capabilities of diffusers on sky, we show 4 hours of differential photometry of 16 Cyg A, taken during our engineering run in September 2016 (see Figures \ref{fig:cygniphotometrypanel}, \ref{fig:3dview}, and \ref{fig:16cygnipanel}). The unbinned raw photometry and unbinned differential photometry (without detrending) are shown in the top and middle panels in Figure \ref{fig:cygniphotometrypanel}, respectively. The unbinned undetrended precision is 776ppm for a 21s cadence. From the top panel, we see that clouds appeared $\sim$1 hour after the observations started (JD$\sim$0.63) and towards the end, but those transparency changes are effectively canceled out in the differential photometry in the middle panel. However, we do observe a slow downward linear trend in the differential photometry, along with a bump at around 0.70 JD with a clear correlation with the $x$ centroid coordinate of the target star (bottom panel).

The diffused PSF remained stable throughout the night. Figure \ref{fig:3dview} shows an image taken at random from this dataset. Speckles are still seen in the diffused PSF, but they are smoothed to a certain extent by seeing variations, and clouds passing by. A video of the observations can be found in the online version of the manuscript. During these observations, the rotator did not work reliably, as it was difficult to rotate the diffuser bearing due to friction. This caused the diffuser to sometimes rotate and sometimes get stuck. Therefore, although the rotator was formally on, the diffuser was only sporadically rotated during the observations presented. This is seen in the video below: the PSF speckle pattern is sometimes smoothed out. Regardless, the overall size and shape of the diffused PSF is not affected. 

Figure \ref{fig:16cygnipanel}a shows the final detrended photometry of 16 Cygni, detrended with a line, airmass, and $x$ and $y$ centroid coordinates. The unbinned photometry (cadence: 21s), shows a precision of 494ppm, while in 30 min bins, the RMS precision bins down to $62^{+26}_{-16}$ppm. The photometric errorbars shown in Figure \ref{fig:16cygnipanel}a are total errors including scintillation, as calculated by Equation \ref{eq:stdtot}. The errorbars increase towards the end due to the increasing airmass. Also shown is the calculated photometric noise error (including photon noise, dark, read, and sky background noise), as calculated by Equation \ref{eq:stdflux} by AstroImageJ, and a representative scintillation error calculated from Equation \ref{eq:stdscint}. The values for these errors in the unbinned photometry are $\sim$100ppm, and $\sim$500ppm, respectively, indicating that these observations are scintillation limited.

To study the effect of residual systematics, we plot the RMS scatter of our 16 Cyg A observations as a function of increasing bin size (Figure \ref{fig:16cygnipanel}b). 
Also overplotted shown are the relative photometric flux errors (green curve) as calculated by AstroImageJ (Equation \ref{eq:phot}), including photon, dark, and read, and sky-background noise, and the expected total photometric noise (including scintillation noise), $\sigma_{\mathrm{tot}} = \sqrt{\sigma_{\mathrm{rel\ flux}}^2 + \sigma_{\mathrm{scint}}^2}$ (blue curve), as calculated using Equations \ref{eq:stds2} and \ref{eq:stdtot}. For the airmass term in Equation \ref{eq:stds2}, we assumed a fixed airmass equal to the mean airmass of the observations, amounting to $\chi=1.10$, which matches well with the observed scatter. We see that our data (black curve) largely bin down as white noise (blue curve).

In Figure \ref{fig:16cygnipanel}b, we also compare our precision to \textit{Kepler}. We ran the 6.5 hour combined differential photometric precision (CDPP) metric, as defined in \cite{gilliland2011} for each of the individual quarters for 16 Cygni A in both long and short cadence mode, that were available from MAST. The resulting 6.5h CDPP precision is consistently at the 7-9ppm level across different quarters, except for Quarter 6, giving a 6.5h CDPP precision of $>$22ppm, which was known to have poor data quality due to the use of a non-optimized photometric aperture \citep{lund2014}. Scaling this result to 30 minutes (i.e.,~$\sigma_{\textrm{CDPP,30min}} = \sigma_{\textrm{CDPP,6.5h}} \sqrt{6.5\textrm{hours} / 0.5 \textrm{hours}}$) gives a precision ranging from 26-30ppm in 30 minute bins across the different quarters, with a median precision of 27.4ppm. Calculating the standard deviation of the long-cadence data ($\sim$30 minute cadence), after throwing out 5 sigma outliers, and running the same 2-day Savitsky-Golay high pass filter as performed in \cite{gilliland2010}, and in the Everest 2.0 pipeline \citep{luger2016,luger2017}, we get a precision of 28-44ppm with a median value of 31.6ppm, in 30 minute bins. This latter way to estimate the precision might be biased towards outliers, but gives a precision estimate that is roughly consistent with the scaled CDPP precision. We choose to use the better of the two, or 27.4ppm as our \textit{Kepler} precision comparison of 16 Cyg A, and we plot this number on Figure \ref{fig:16cygnipanel}b. Additionally, in Figure \ref{fig:16cygnipanel}b, to graphically compare an equal length segment of \textit{Kepler} photometry to our 4 hour ground-based photometry, we plot a representative 4 hour segment of \textit{Kepler} short cadence data of 16 Cyg A. We say that this 4 hour segment is "representative" of \textit{Kepler} photometry on this star, as in 30 minute bins the precision of this 4 hour segment is similar to the 27.4ppm 30-minute CDPP precision value discussed above. Additionally, to perform a head-to-head comparison with our ground-based differential photometry to \textit{Kepler's} differential photometry of 16 Cyg A, we calculated the same 30 minute CDPP metric for Quarter 7 on the \textit{differential} 16 Cyg A short-cadence \textit{Kepler} light-curve, using 16 Cyg B as a reference star. In doing this, the resulting 30 min CDPP metric for the 16 Cyg Quarter 7 differential light curve degraded to 36.3ppm, effectively adding a factor of $\sqrt{2}$ to the photometric noise. By this metric, our ground-based precision of 16 Cyg A of $62^{+26}_{-16}$ppm in 30 minute bins is a factor of $<2$ from \textit{Kepler}'s differential precision (37.4ppm), but a factor of $\sim$2 in the non-differential photometry case (27.4ppm). We show the exact 4 hour segments of \textit{Kepler} photometry of 16 Cyg A and B used in Figure \ref{fig:kepler_panel}a in the Appendix.

\begin{figure*}[t]
	\begin{center}
		\includegraphics[width=0.9\textwidth]{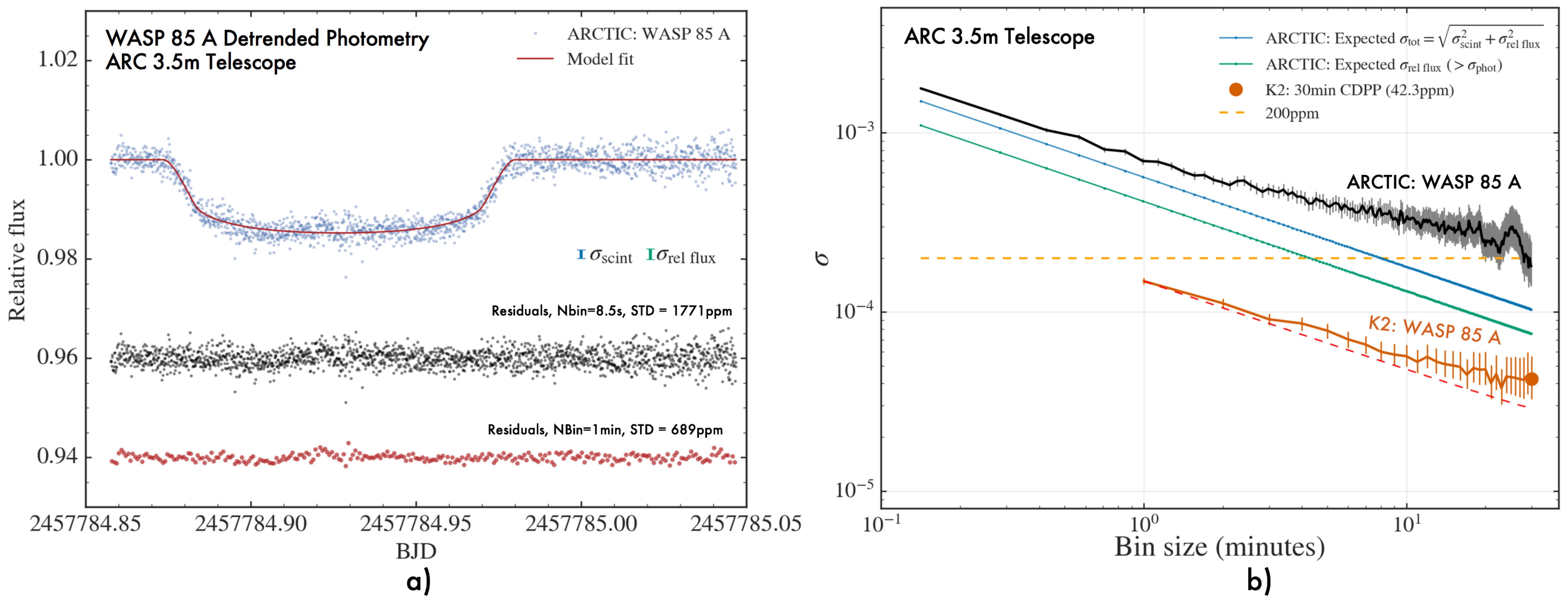}
	\end{center}
	\caption{a) Transit of WASP 85 A b observed with the diffuser on ARCTIC, along with a best fit model, and residuals offset for clarity. The unbinned precision (black points) is 1771ppm, while binning to 1 minute, gives a precision of 689ppm (red points). We attribute the "bump" around the middle of the transit to the planet crossing a starspot; b) Photometric precision comparison with \textit{K2}: photometric precision as a function of bin size in minutes for our 4.5 hours of ground-based diffuser-assisted transit observations of WASP 85 A b (black curve), as compared to a representative 4.5 hour \textit{K2} short-cadence photometry of the same star out of transit (orange curve; see text and Figure \ref{fig:kepler_panel} for details). Additionally shown is the relative photometric error, $\sigma_{\mathrm{rel\ flux}}$ (green curve) for our diffuser observations, along with the total expected error $\sigma_{\mathrm{tot}}$ (blue curve). The diffuser assisted observations reach $180_{-41}^{+66}$ppm in 30 minutes. A machine readable table including the photometry in panel a) is available in the online journal.}
	\label{fig:wasp85panel}
\end{figure*}

\paragraph{Transit of WASP 85 A b} 
Figure \ref{fig:wasp85panel}a shows the transit of WASP 85 A b as observed with ARCTIC, along with our best fit transit model after detrending the lightcurve with $x$ and $y$ pixel coordinate, airmass and a line. The unbinned residuals are shown offset for clarity, showing a photometric precision of 1771ppm. Also shown is the residuals binned to 1 minute bins, with a precision of 689ppm. Additionally shown in the plot are representative scintillation, and a photon noise error bars. 

We observe a small brightening---or ``bump''---in the middle of the transit. We speculate this is caused by the planet crossing a starspot, as there is no clear correlation with other observational parameters, suggesting the bump is of astrophysical origin. This is a likely scenario, as WASP 85 A has been observed to have repeated starspot crossing events, through detailed analysis of \textit{K2} data \citep{mocnik2016}. If the bump in Figure \ref{fig:wasp85panel}a is indeed a starspot crossing, this marks the first ground-based detection of a starspot crossing event for this target.

Figure \ref{fig:wasp85panel}b, shows how the RMS scatter of the WASP 85 A b residuals change with increasing bin size. Also shown are the expected $\sigma_{\mathrm{rel\ flux}}$ (green curve), and total errors $\sigma_{\mathrm{tot}} = \sqrt{\sigma_{\mathrm{rel\ flux}}^2 + \sigma_{\mathrm{scint}}^2}$ (blue curve), as calculated by Equations \ref{eq:stdflux} and \ref{eq:stdtot}, respectively. The scintillation noise was calculated using Equation \ref{eq:stdscint} assuming three reference stars, and using the mean airmass of the observations of 1.3. Different from our 16 Cygni observations, where $\sigma_{\mathrm{rel\ flux}} \ll \sigma_{\mathrm{scint}}$, these observations are in the regime where $\sigma_{\mathrm{rel\ flux}} \sim \sigma_{\mathrm{scint}}$. We do observe that our obtained unbinned precision (1771ppm) is slightly higher than the expected (1500ppm). This might be attributed to some degree to the uncertainty in the estimation of the scintillation noise, which has a strong dependence on the variable wind profile in the upper atmosphere \citep{osborn2015}. An additional factor helping in explaining this discrepancy, is that the three reference stars are all much fainter than the target star (having fluxes of 3\%, 10\%, and 3\% of the target star, respectively), lowering their S/N in correcting for atmospheric transparency fluctuations. Lastly, some of the correlated noise behavior towards higher bin sizes in Figure \ref{fig:wasp85panel}b, might be explained by astrophysical noise, including the candidate starspot crossing event.

Similar to Figure \ref{fig:16cygnipanel}, we compare our achieved precision on WASP 85 A, with the 30-minute CDPP precision achieved by \textit{K2} in Figure \ref{fig:wasp85panel}b. The \textit{K2} short-cadence photometry of WASP 85 A from Campaign 1, was retrieved and detrended with the Everest 2.0 pipeline \citep{luger2016,luger2017}. The 6.5hour CDPP precision of \textit{K2} as calculated by Everest is 11.7ppm, and using a similar scaling as in the discussion above to calculate the 30 minute CDPP precision (i.e.,~ $\sigma_{\textrm{CDPP,30min}} = \sigma_{\textrm{CDPP,6.5h}} \sqrt{6.5\textrm{hours} / 0.5 \textrm{hours}}$), yields a 30 minute CDPP of 42.3ppm. Similarly to the discussion above for 16 Cyg A in Figure \ref{fig:16cygnipanel}b, in Figure \ref{fig:wasp85panel}b we plot the $\sigma$-vs-bin-size curve for a representative 4.5 hour segment of \textit{K2} short cadence data, the same length as our observations with ARCTIC. The specific segment is shown in Figure \ref{fig:kepler_panel}b in the Appendix. We see that on this star, our 30-minute ground-based precision ($180^{+66}_{-41}$ppm) is a factor of $\sim$4 to that of the \textit{K2} observations on this star (42.3ppm).

\begin{figure*}[t]
	\begin{center}
		\includegraphics[width=0.9\textwidth]{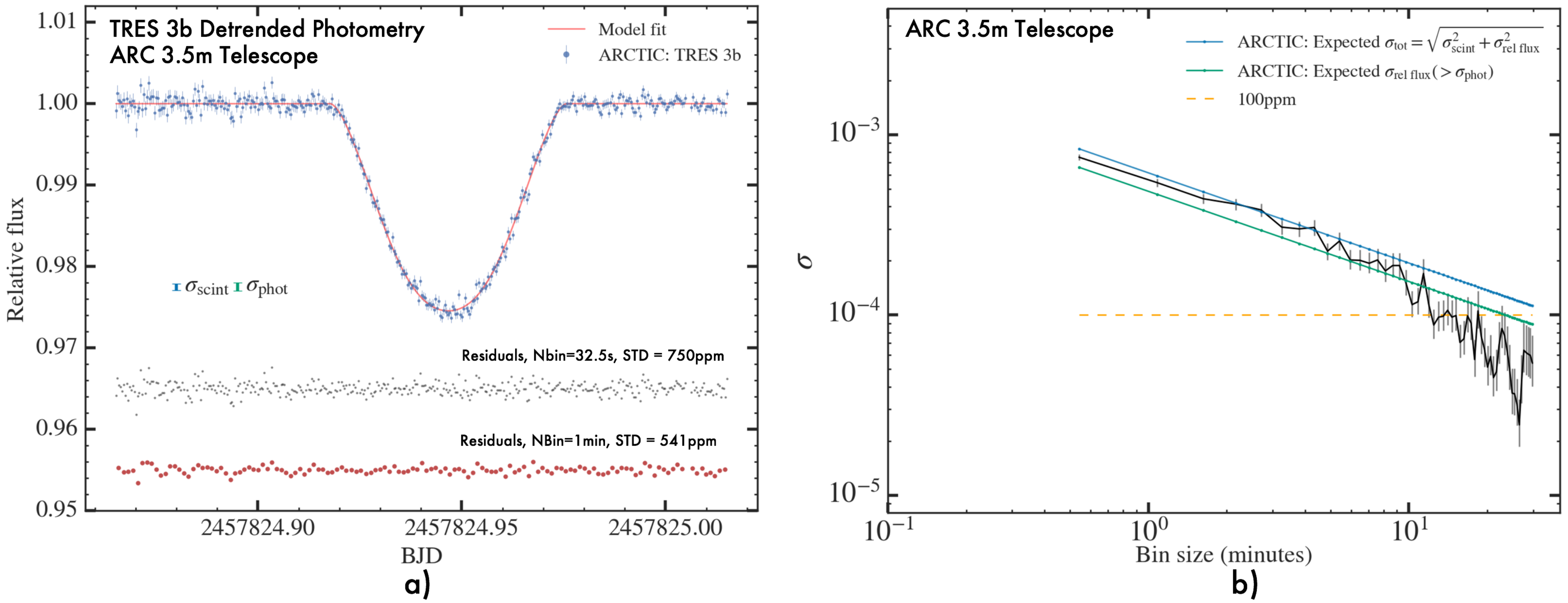}
	\end{center}
	\caption{a) Transit of TRES 3 b as observed with the diffuser on ARCTIC, along with a best fit model, and residuals offset for clarity. The unbinned precision (black residuals) is 750ppm, while binning to 1 minute, gives a precision of 541ppm (red residuals). b) Photometric precision as a function of bin for the data shown in the left panel (black curve). Additionally shown is the relative photometric error, $\sigma_{\mathrm{rel\ flux}}$ (green curve) for our diffuser observations, along with the total expected error $\sigma_{\mathrm{tot}}$ (blue curve). The diffuser assisted observations reach $54^{+23}_{-14}$ppm in 30 minutes, which is below the expected photometric precision (blue curve) of 101ppm. We conservatively say that we reach a precision of 101ppm in 30 minute bins. A machine readable table including the photometry in panel a) is available in the online journal.}
	\label{fig:trespanel}
\end{figure*}

We note that our precision of $180_{-41}^{+66}$ppm in 30 minute bins for our WASP 85 A b transit observations, is worse than the precision for our 16 Cyg A observations. This is due to a few reasons. First, 16 Cyg A and B are both brighter than WASP 85 A and the available reference stars in the WASP 85 A field. This allowed us to suppress the photon noise better for the 16 Cyg A observations, while the WASP 85 A b transit observations are in the regime where the photon noise and scintillation noise are similar. Additionally, 16 Cyg A and B are spectrally well matched, reducing the effects from secondary extinction. Second, for the 16 Cyg A observations, we used a filter (Semrock 857/30) with little-to-no water absorption in the red-optical, minimizing systematics due to molecular absorption and extinction. Furthermore, astrophysical systematics for WASP 85 A, in particular from our candidate starspot crossing event, further add to the systematic noise floor for those observations. Still, even though the precision level achieved for our WASP 85 A transit observations is worse than for our 16 Cygni observations, the precision of $180_{-41}^{+66}$ppm in 30 minutes demonstrates that diffusers are enabling precision observations across a wide magnitude range, even for targets with overlapping PSFs.

\paragraph{Transit of TRES 3 b}
Figure \ref{fig:trespanel}a shows our in-transit observations of TRES-3b as observed by ARCTIC with a diffuser, along with our best fit transit model after detrending the lightcurve with $x$ and $y$ pixel centroids, and airmass. We did not detrend with a straight line as we did for the other observations, as it did not yield a significant improvement in the RMS scatter.

Figure \ref{fig:trespanel}b shows how the RMS scatter of the TRES 3b residuals change with increasing bin size, similar to Figures \ref{fig:cdk24phot}b, \ref{fig:16cygnipanel}b, and \ref{fig:wasp85panel}b. 
Furthermore, Figure \ref{fig:trespanel}b compares the relative photometric flux errors, $\sigma_{\mathrm{rel\ flux}}$ (green curve), to the expected total photometric noise (including scintillation noise), $\sigma_{\mathrm{tot}} = \sqrt{\sigma_{\mathrm{rel\ flux}}^2 + \sigma_{\mathrm{scint}}^2}$ (blue curve), as calculated using Equations \ref{eq:stdflux}, and \ref{eq:stdtot}, respectively.
For the scintillation term in Equation \ref{eq:stds2}, we assumed a fixed airmass equal to the mean airmass of the observations ($\chi=1.20$), and 10 reference stars. For the smallest bins, we see that the scatter bins down roughly as white noise. However, at the largest bins, we see a dip in the precision below the expected Gaussian behavior at bin sizes >10minutes: the data bin down to $54^{+23}_{-14}$ppm in 30 minute bins (black curve), slightly below the expected photometric precision of 101ppm (blue curve). The drop in Figure \ref{fig:trespanel}b is somewhat below the expected photometric noise (green curve in Figure \ref{fig:trespanel}b), but is overall largely within the 1 or 2-sigma errorbars. Similar excursions below the Gaussian expected behavior at large bin sizes have been reported in the literature (e.g., \cite{cubillos2013}, and \cite{blecic2013}), and \cite{cubillos2017} demonstrate that those fluctuations are not statistically significant after taking into account the increasingly skewed inverse Gamma distribution of the bins at large bins sizes. Therefore, we argue that a precision much below the expected photometric noise (blue curve) is likely an overestimate of the actual precision, and we thus conservatively say that our achieved precision in 30 minutes for these observations equals the expected photometric precision of 101ppm. We discuss this further in Section 6.1, where we compare our precision with other values reported in the literature.

\subsection{NIR diffuser-assisted photometry}
\label{sec:nirresult}
\begin{figure*}
	\begin{center}
		\includegraphics[width=0.9\textwidth]{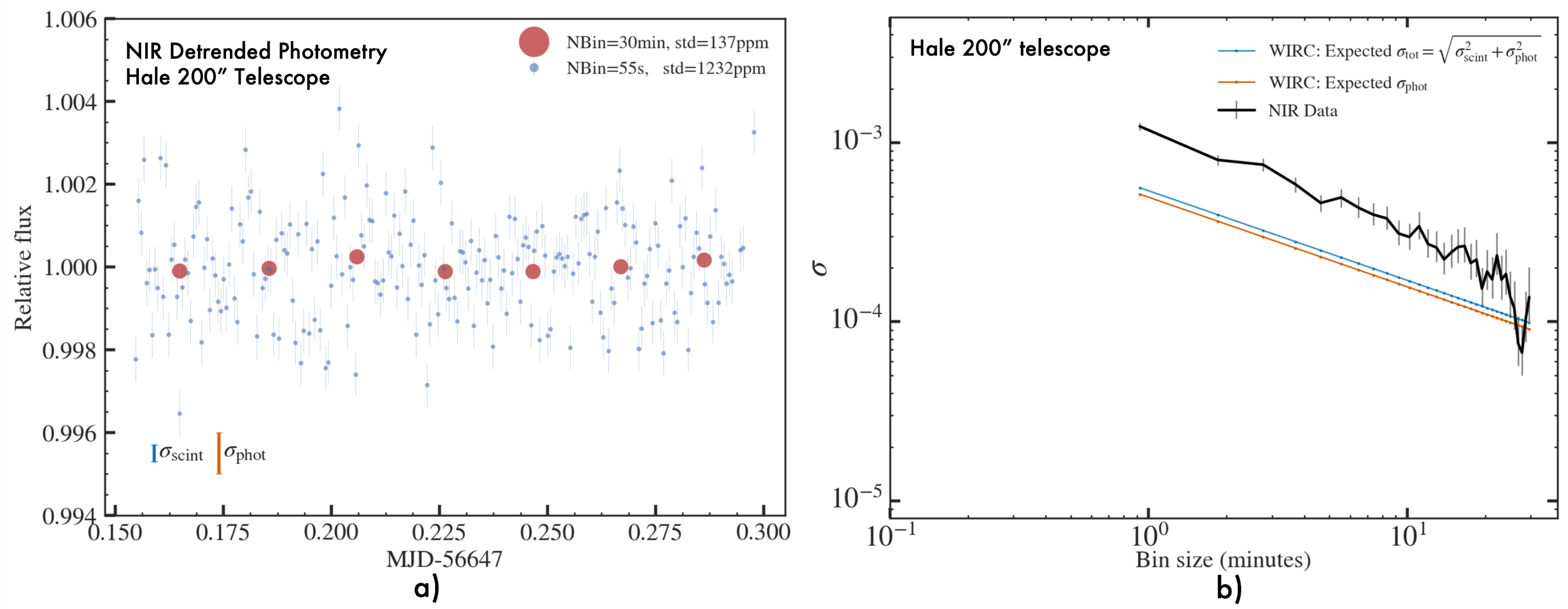}
        \vspace{-0.5cm}
	\end{center}
	\caption{Results from the on-sky diffuser test in the NIR at Palomar. a) Final unbinned photometry (blue points) shown with total expected errors including scintillation errors, along with the photometry in 30 minute bins (red points). b) Photometric precision as a function of bin size in minutes (black curve), showing that the noise in the data is largely white. The orange curve shows the expected photon noise (without read, dark, or background noise). The blue curve shows the scintillation noise added in quadrature to the photon noise. In 30 minute bins the data bin down to $137^{+64}_{-36}$ppm. A machine readable table including the photometry in panel a) is available in the online journal.}
	\label{fig:nir_diffuser_result}
\end{figure*}

Figure \ref{fig:nir_diffuser_result} shows our initial test results with the diffuser in the NIR, showing the reduced light curve of the target star 2MASS J04230271+5740319 after normalizing with the 10 reference stars used and the median flux. The original scatter of the light curve is 1232ppm before any binning. 

Figure \ref{fig:nir_diffuser_result}b shows how the scatter of the binned light curve changes with increasing bin size, where we also show the expected photon noise (excluding background noise) for these observations in orange, along with the scintillation noise as calculated by Equation \ref{eq:stdtot}, added in quadrature to the photon noise. We see that the scatter of the binned light curve follows largely the expected binning of Gaussian white noise, and similar to the TRES 3 observations above (Figure \ref{fig:trespanel}b), we observe an excursion below the expected photon noise (orange curve) at the highest bin sizes. Similar to the TRES 3 observations, we attribute this behavior to the small number of bins at the higher bin sizes. The data (black curve) bins down to $137^{+64}_{-36}$ppm in 30 minute bins, which is above the expected total noise of $\sigma_{\mathrm{tot}} \sim 100$ppm. These precision levels of $137^{+64}_{-36}$ppm in 30 minute bins, are among the best broad-band photometric precisions achieved in the NIR from the ground. 

For these observations, the expected scintillation is small due to the large aperture of the telescope, the large number of reference stars used, along with the low airmasses of the observations. We however, note the important caveat that the total error, $\sigma_{\mathrm{tot}}$ (blue curve), does not include a term from the background noise, as the previous figures have, which will be particularly important in the NIR. We did not calculate the background noise term here, as these observations used a dedicated dithered sky-background master frame subtracted from the raw images to create the final science frames, different from our observations in the optical.

\section{Transit MCMC fitting}
\label{sec:mcmc}

To check that our planet parameters agree with the values reported in the literature, we performed Markov Chain Monte Carlo (MCMC) fitting to the light curves of WASP 85 A b, and TRES-3b. We performed our MCMC fitting following a similar $\chi^2$-minimization modeling approach as the EXOFAST code as described in \cite{Eastman2013}. Different from EXOFAST, our code is in Python, which utilizes the \texttt{BATMAN} Python package \citep{kreidberg2015}, which uses the \cite{mandel2002} transit model formalism. Similar to EXOFAST, we first use an amoeba minimization algorithm to find a good $\chi^2$ minimum in the parameter space describing the transit. We use those parameter values as starting points for our MCMC chains, initializing 30 walkers for the \texttt{emcee} affine-invariant ensamble sampler \citep{dfm2013}, in a small Gaussian ball centered around the amoeba best-fit values.

For our MCMC chains, we used 5 jump parameters describing the planet transit: $T_0$, $\log(P)$, $\cos(i)$, $R_p/R_*$, $\log(a/R_*)$, along with one parameter for the out-of-transit baseline flux. Following the modeling efforts of the discovery papers for WASP 85 A b \citep{brown2014}, and TRES-3b \citep{odonovan2007}, we fix the eccentricity and the argument of periastron to be equal to 0. To account for systematic correlated noise in our data, we include up to 4 additional detrending coefficients in the $\chi^2$-minimization including: the normalized airmass, $x$ and $y$ centroid pixel coordinates, and a straight line through the data. For the WASP 85 A b we use all 4 detrending parameters, but for TRES-3b we use the first three detrending parameters, as detrending with a line did not improve the quality of the fit. 

Our priors are summarized in Table \ref{tab:priors}. For WASP 85 A b, we adopt Gaussian priors on the planet parameters from \citep{brown2014}, which we propagated accordingly using standard error propagation to describe our modified MCMC jump parameters (Table \ref{tab:priors}). For TRES-3b, we adopt Gaussian priors on the stellar and planet parameters from \citep{sozzetti2009}, but we adopt our priors on the transit ephemeris from the more recent work of \cite{jiang2013}. For both transits, we kept the limb-darkening parameters fixed, calculating their values using a quadratic limb-darkening law from \cite{claret2011} for the respective band-passes using the EXOFAST limb darkening web applet\footnote{\url{http://astroutils.astronomy.ohio-state.edu/exofast/limbdark.shtml}}, which uses the host star $\log g$, $T_{\mathrm{eff}}$, and [Fe/H] values. To account for the light-curve flux dilution from the the WASP 85 B binary companion in the WASP 85 A b light curve, we follow the approach described in \citep{brown2014}, adopting a fixed dilution factor for third-light in our observing band of $L_{\mathrm{3,SDSS i^\prime}} = 0.5$ to the $R_p/R_*$ ratio. For the detrending parameters, we chose Gaussian priors on the best-fit values found by the initial amoeba $\chi^2$-minimization.

After initializing each MCMC walker in \texttt{emcee}, we ran each walker for 5000 steps and threw the first 1000 steps out as burn-in. Two dimensional corner plots of the resulting chains are shown in Figures \ref{fig:wasp_corner}, and \ref{fig:tres_corner} in the Appendix. Furthermore, to check if the chains were ready for inference, we followed the suggestion by \cite{ford2006} to check the Gelman-Rubin statistic, $\hat{R_{\nu}}$, which was within 5\% of unity for all of the parameters. The mean acceptance fraction for in the sampling of the MCMC chains were $\sim$43\%, $\sim$44\%, for the WASP 85 A b, and TRES 3b light curves, respectively. The results from our MCMC fits are listed in Table \ref{tab:tresparams}, showing the median best-fit planet parameters, along with 68\% confidence intervals. 

For WASP 85 A b, our overall planet parameters agree well with the values reported in \cite{brown2014}, including our best-fit planet radius of $1.515_{-0.043}^{+0.044} R_J$.
However, both our value and the value reported by \citep{brown2014} ($R_p = 1.48 \pm 0.03 R_J$) are somewhat higher than the value reported by \cite{mocnik2016}. We speculate that this discrepancy is due to systematics arising from observing the target in different filters---with the work presented here, and in \cite{brown2014} done in red-optical filters, and the work by \cite{mocnik2016} in the blue \textit{Kepler} bandpass---due to contaminating light from the WASP 85 B binary companion. Furthermore, we note that our median value for the orbital inclination of $88^{\circ}.89_{-1.1}^{+0.77}$ is somewhat lower than the inclination reported by \cite{brown2014} ($89^{\circ}.72_{-0.24}^{+0.18}$) and \cite{mocnik2016} ($89^{\circ}.69_{-0.03}^{+0.11}$). Looking at our $\cos(i)$ posteriors in Figure \ref{fig:wasp_corner} in the Appendix, we see that our $\cos(i)$ inclination distribution does not approximate a Gaussian profile. We attribute this to the $\cos(i)$ inclination distribution being a positive-definite parameter (as we constrain $i<90^{\circ}$), causing the median of the $\cos(i)$ distribution to be biased towards positive values, although the mode of the distribution is close to or consistent with 0---similar to the \cite{lucysweeney1971} eccentricity bias.

\begin{deluxetable*}{llcc}
\tablecaption{Summary of priors for TRES-3b and WASP 85 A b. Priors for the MCMC values are Gaussian priors adopted from \cite{brown2014} for the WASP 85 A b stellar and transit parameters. For our TRES 3b light curve, we adopt priors from \cite{sozzetti2009} on the stellar parameters, and more recent parameters from \cite{jiang2013} on the transit ephemeris.\label{tab:priors}}
\tablehead{\colhead{~~~Parameter}                                 &  \colhead{Description}                  & \colhead{WASP 85 A b}        & \colhead{TRES-3b}}
\startdata
\multicolumn{4}{l}{\hspace{-0.2cm} Stellar Parameters:}           \\
$R_* (R_\odot)$                                                   &  Stellar radius                         & $0.935\pm0.023$              & $0.829\pm0.022$              \\
$T_{\mathrm{eff}}$ (K)                                            &  Stellar effective temperature          & $5685\pm65$                  & $5650\pm75$                  \\
$\mathrm{[Fe/H]}$                                                 &  Stellar metallicity                    & 0.08                         & -0.19                        \\
$\log(g)$                                                         &  Stellar gravity                        & 4.519                        & 4.568                        \\
\multicolumn{4}{l}{\hspace{-0.3cm} MCMC transit jump parameters:} \\
$T_{C}$ $(\mathrm{BJD_{TDB}})$                                    &  Transit Midpoint                       & $2457784.927 \pm 0.00020$    & $2457824.94589 \pm 0.00006$  \\
$\log(P)$ (days)                                                  &  Orbital period                         & $0.424175367 \pm 0.00000007$ & $0.116005357 \pm 0.00000005$ \\
$\cos(i)$                                                         &  Transit inclination                    & $0.0054 \pm 0.0019$          & $0.1418 \pm 0.0028$          \\
$R_p/R_*$                                                         &  Radius ratio                           & $0.1358 \pm 0.0046$          & $0.1653 \pm 0.0063$          \\
$\log(a/R_*)$                                                     &  Normalized orbital radius              & $0.951 \pm 0.011$            & $0.7724 \pm 0.0076$          \\
\multicolumn{4}{l}{\hspace{-0.3cm} MCMC fixed parameters:}        \\
$\sqrt{e} \cos(\omega) $                                          &  Eccentricity \& Argument of periastron & 0 (adopted)                  & 0 (adopted)                  \\
$\sqrt{e} \sin(\omega) $                                          &  Eccentricity \& Argument of periastron & 0 (adopted)                  & 0 (adopted)                  \\
$u_1$                                                             &  Linear limb-darkening coefficient      & 0.4122                       & 0.3145                       \\
$u_2$                                                             &  Quadratic limb-darkening coefficient   & 0.2680                       & 0.2704                       \\
$L_{3,\mathrm{SDSS r^\prime}}$                                    &  Dilution ratio                         & 0.50                         & \nodata                      \\
\enddata
\end{deluxetable*}

\begin{deluxetable*}{llcc}
\tablecaption{Median values and 68\% confidence intervals for the transit fit parameters for WASP 85 A b, and TRES 3b.\label{tab:tresparams}}
\tablehead{ \colhead{~~~Parameter}& \colhead{Description}  & \colhead{WASP 85 A b}                  & \colhead{TRES 3 b}}
\startdata
$T_{C}$ $(\mathrm{BJD_{TDB}})$ & Transit Midpoint          &   $2457784.92695_{-0.00017}^{+0.00017}$&$2457824.946728_{-0.000008}^{+0.000009}$     \\
$P$ (days)                     & Orbital period            &   $2.6556777_{-0.0000004}^{+0.0000004}$ &  $1.3061870_{-0.0000001}^{+0.0000001}$     \\
$R_p/R_*$                      & Radius ratio              &            $0.1666_{-0.0024}^{+0.0025}$ &        $0.17079_{-0.00037}^{+0.00039}$     \\
$R_p (R_\oplus)$               & Planet radius             &                 $16.99_{-0.48}^{+0.50}$ &                $15.44_{-0.41}^{+0.41}$     \\
$R_p (R_J)$                    & Planet radius             &               $1.515_{-0.043}^{+0.044}$ &              $1.377_{-0.036}^{+0.037}$     \\
$\delta$                       & Transit depth             &         $0.02775_{-0.00080}^{+0.00085}$ &        $0.02917_{-0.00013}^{+0.00013}$     \\
$a/R_*$                        & Normalized orbital radius &                  $8.71_{-0.33}^{+0.14}$ &           $5.8991_{-0.0079}^{+0.0077}$     \\
$a$ (AU)                       & Semi-major axis           &            $0.0376_{-0.0015}^{+0.0012}$ &        $0.02273_{-0.00060}^{+0.00060}$     \\
$i$ $(^{\circ})$               & Transit inclination       &                 $88.89_{-1.10}^{+0.77}$ &             $81.683_{-0.021}^{+0.021}$     \\
$b$                            & Impact parameter          &                  $0.17_{-0.12}^{+0.15}$ &           $0.8533_{-0.0010}^{+0.0011}$     \\
$e$                            & Eccentricity              &                             0 (adopted) &                            0 (adopted)     \\
$\omega$ $(^{\circ})$          & Argument of periastron    &                             0 (adopted) &                            0 (adopted)     \\
$T_{\mathrm{eq}}$(K)           & Equilibrium temperature   &                      $1366_{-21}^{+28}$ &                     $1645_{-22}^{+22}$     \\
$T_{14}$ (days)                & Transit duration          &            $0.1124_{-0.0011}^{+0.0016}$ &        $0.05728_{-0.00004}^{+0.00004}$     \\
$\tau$ (days)                  & Ingress/egress duration   &            $0.0165_{-0.0006}^{+0.0016}$ &        $0.02864_{-0.00002}^{+0.00002}$     \\
$T_{S}$ $(\mathrm{BJD_{TDB}})$ & Time of secondary eclipse &         $2457786.25479_{-0.00017}^{+0.00017}$ &    $2457825.5998215_{-0.000008}^{+0.000009}$     \\
\enddata
\end{deluxetable*}

For TRES-3b, our overall planet parameters agree well with the values reported by \cite{sozzetti2009}. We note that our best-fit planet radius of $1.377_{-0.036}^{+0.037} R_J$, is slightly higher than that the value of $1.336_{-0.037}^{+0.031} R_J$ in \cite{sozzetti2009}. Acknowledging that $R_p/R_*$ can be sensitive to the exact detrending parameters chosen for ground-based observations \citep[e.g.,][]{villanueva2016}, and as our value is within 2-$\sigma$ from the values reported by \cite{sozzetti2009}, we conclude that our value is consistent with their result. Assuming the ephemeris from \cite{jiang2013} for TRES-3b, our transit midpoint of $T_C = 2457824.946728_{-0.000008}^{+0.000009}$ differs only 1 minute from the expected linear ephemeris, demonstrating that our transit ephemeris shows no evidence for significant transit timing variations.

\section{Discussion}
\subsection{A comparison with other high-precision photometry}
\label{sec:precisioncomparison}
We compare our photometric precision to some of the highest ground-based precisions reported in the literature in Table \ref{tab:precision}, in the optical and NIR. The methods used in Table \ref{tab:precision} to achieve these photometric precisions include telescope defocusing \citep{gilliland1993,kundurthy2013,tregloan2013,fukui2016,croll2011,zhao2014}, using orthogonal-transfer CCDs \citep{johnson2009}, tunable filter narrow-band spectrophotometry \citep{Colon2012}, or diffusers (this work). Our goal here is to survey some of the best published photometric precision values from the ground, and to show that the first diffuser-assisted photometric precisions are already paralleling the best precisions in the literature.

The most direct way to compare the photometric precision levels achieved across these studies is to compare the precision per unit time. Another comparison metric is to compare how close the achieved precision to the expected photon and scintillation limit, as we have done in Figures \ref{fig:cdk24phot}, \ref{fig:16cygnipanel}, \ref{fig:wasp85panel}, \ref{fig:trespanel} and \ref{fig:nir_diffuser_result}. However, this is often challenging to do for other efforts in the literature without knowledge of the telescope throughput, and/or with the image frames in hand, as the scintillation and photon noise errors are often not reported. Due to these uncertainties, we restrict our comparison in Table \ref{tab:precision} to comparing reported photometric precision levels in the respective papers per unit time, and only compare them to the photon noise limit when explicitly stated in the paper.

For this comparison, we choose to use two metrics. First, we compare the precision levels per one minute effective cadence, $\sigma_{\textrm{1min}}$, a regime where correlated noise in practice is a minimal fraction of the overall photometric noise budget. To acquire the $\sigma_{\textrm{1min}}$ values, we either adapt it directly as mentioned in the paper, or if the precision in 1 minute is not specifically reported, we calculate a scaled value from the unbinned precision assuming pure Gaussian white noise behavior, i.e.,~using $\sigma_{\textrm{1min}} = \sigma_{\textrm{unbinned}} / \sqrt{60s/t_{\textrm{cadence}}}$, where $t_{\textrm{cadence}}$ is the effective cadence of the observations (time between successive exposures). Second, we also choose to compare the reported precision levels at 30 minute bins as $\sigma_{\textrm{30min}}$. 
This is the binning regime where correlated noise can be a significant fraction of the overall error. The level of correlated noise is often not clear without the light-curve in hand; therefore, to estimate these values, we did either of the following. We either report the $\sigma_{\textrm{30min}}$ value in Table \ref{tab:precision} if we can specifically read the value from the paper (e.g.,~from a precision-vs-bin plot), or, if the $\sigma_{\textrm{30min}}$ value was not clearly visible from the paper, we calculated $\sigma_{\textrm{30min}}$ by performing a best-fit transit fit using the published light-curve datapoints from the the paper. In the ensuing discussion, we choose to keep the comparison between the optical and NIR photometric precisions separate, due to the dissimilar systematics and instrument technologies involved.

Using the $\sigma_{\textrm{1min}}$ metric, in the optical, we see that the 16 Cygni observations presented here with $\sigma_{\textrm{1min}} = 300$ppm, are the overall highest precision, with other efforts coming very close. Notably, these include \cite{kundurthy2013}, achieving a precision of $\sigma_{\textrm{1min}} = 306$ppm through telescope defocusing also on the 3.5m telescope at APO. We discuss the \cite{kundurthy2013} observations further below as a comparison between defocused and diffuser-assisted observations.

Using the $\sigma_{\textrm{30min}}$ metric, we see that diffuser-assisted precision levels are already matching some of the best published photometric precisions presented in the literature. Notably, these efforts include \cite{Colon2012} achieving $\sigma_{\textrm{30min}}=\sim$65ppm using tunable-filter narrow-band spectrophotometry on the 10.4m GTC telescope\footnote{Value read from the best precision-vs-bin plot presented in \cite{Colon2012}}, and \cite{johnson2009}, achieving $\sigma_{\textrm{30min}}=72^{+26}_{-16}$ppm precision using the Orthogonal Parallel Transfer Imaging Camera on the 2.2m telescope at Maunakea\footnote{Value calculated by fitting a best-fit transit to the published light-curve data from \cite{johnson2009}}, and \cite{fukui2016}, achieving $24^{14}_{-7}$ppm precision using the MuSCAT imager on the 1.88m at the Okayama Astrophysical Observatory in Japan\footnote{Value calculated by fitting a best-fit transit to the published light-curve data from \cite{fukui2016}}. For \cite{Colon2012} and \cite{johnson2009} to achieve photometric precision levels similar to the precisions achieved in this work is not unexpected for a few reasons. First, their efforts are a specialized implementation of the general techniques that we promote (\cite{Colon2012} using narrow band filters to reduce systematics and spreading out the light over many pixels; and \citep{johnson2009} deterministically molding the PSF to a broad and stable shape throughout the observations). Taking into account that scintillation noise is further averaged out with larger telescope diameters as $D^{-2/3}$ (see Equation \ref{eq:stds2}), further places these observations at very similar values.

In doing this analysis we noted that some of the measurements took large excursions down well below the expected Gaussian white noise limit---similar to our TRES-3b observations (achieving formally $54^{+23}_{-14}$ppm precision in 30 minute bins, but with an expected photon-limit value of 101ppm; see Figure \ref{fig:trespanel}). This happened notably for \citep{Colon2012} achieving $\sim$65ppm precision in 30 minute bins, with a Gaussian expected value of 93ppm, and \citep{fukui2016}, achieving formally $24^{14}_{-7}$ppm precision in 30 minute bins, with a Gaussian expected value of 80ppm. Similar to our discussion above in Section \ref{sec:opticalresults} for our TRES-3b observations, we argue that these values are likely an overestimate of the actual precision due to binning induced fluctuations at larger bin sizes, and we suggest that the Gaussian expected values are likely a better estimate of the actual achievable precision. We include both values in Table \ref{tab:precision} for completeness.

In the NIR, we compare the photometric precision achieved here, to two other high-precision NIR photometry efforts in the literature. First, we compare our observations to the defocused observations performed by \cite{croll2011}, observing the secondary eclipse of WASP 12 in the $J$, $H$, and $K_S$ bands, and second, to the defocused observations performed by \cite{zhao2014}, observing the secondary eclipse of HAT-P-32Ab in the $H$ and $K$ bands. To perform a head to head comparison with the \cite{croll2011} and \cite{zhao2014} results, we specifically compare our photometric precision to their $K_S$ observations. By the $\sigma_{\textrm{1min}}$ metric, \cite{croll2011} achieve a better precision than our precision in the NIR, but by the $\sigma_{\textrm{30min}}$ metric, we see that the precision of our NIR diffuser-assisted observations is better. These early diffuser-assisted observations with WIRC are thus suggestive that the diffuser on WIRC is enabling a reliable path to perform routine high-precision observations in the future.

To perform a photometric precision comparison between the diffuser-assisted method and the defocusing method, perhaps the most direct comparison between the two is between our 16 Cyg A photometry and the work by \cite{kundurthy2013}, which performed defocused observations using the AGILE instrument \citep{mukadam2011} also on the 3.5m ARC telescope at APO. Such a comparison normalizes the telescope size and observing site out of the equation, but there are still differences in the target observed, and the instrument setups (e.g.,~AGILE allowed for 100\% efficiency). In the NIR, we can make a similar comparison between the work presented here, and the work by \cite{zhao2014}, performing defocused observations also using the WIRC instrument on the 200" Hale telescope at Palomar. Acknowledging uncertainties in different observing conditions and other technical factors, overall, Table \ref{tab:precision} suggests that diffuser-assisted observations can match or exceed defocused observations with specialized instruments. The main benefit with defocused observations is that the size of the PSF FWHM can be tuned to the observing conditions, but with the downside of being susceptible to PSF variations throughout the night. However, with diffusers and their robust PSF stabilization capabilities, we argue that we can more consistently reach precision levels such as reported in Table \ref{tab:precision}, even in less-than-optimal observing conditions (such as our 16 Cyg observations), assuming the availability of a good reference star.

Of interest for future TESS follow-up, we note that that our demonstrated precision in the optical with ARCTIC in 30minutes matches or surpasses the expected precision of TESS, across the different magnitudes observed. For bright stars ($I_C < 7$) TESS is expected to be limited by systematic noise sources at the $\sim$60ppm level in 1 hour \citep{sullivan2015}. Assuming a white noise scaling, this corresponds to $\sim$84ppm in 30 minutes. Our observations of 16 Cyg A, demonstrating $62^{+26}_{-16}$ppm in 30 minutes (Figure \ref{fig:16cygnipanel}) on an I=5.1 star, are thus at a similar precision level. Second, as discussed by \cite{sullivan2015}, the expected precision of TESS around a $I_C=10$ magnitude star---a brightness similar to WASP 85 A, and TRES-3b---is $\sim$200ppm in 1 hour. Likewise, assuming a white noise scaling, this precision corresponds to $200 \sqrt{2} =$$\sim$282ppm in 30 minutes. With our WASP 85 A b, and TRES-3b transit observations, we demonstrate a 30-minute precision better than this by a factor of 1.6, and 2.8, respectively. This demonstrates that ARCTIC with a diffuser will be capable of following up TESS targets with TESS-like precision across a large range of magnitudes, assuming the availability of good reference stars. We similarly expect that the WIRC instrument with a diffuser will be a work-horse instrument in following up with TESS targets at high-precision. However, with WIRC operating in the NIR, the comparison with the expected TESS precision levels---which are in the optical---is not strictly analogous to comparing to our ARCTIC optical precision levels to TESS, as e.g., the photon count levels for a $K_S=10$ magnitude stars is different from a $I_C=10$ magnitude star.


To reach photometric precision levels to those reported in Table \ref{tab:precision}, requires a judicious attention to minimizing all sources of photometric noise. For the brightest stars (see Figure \ref{fig:scint}) scintillation noise is larger than the photon noise across different telescope systems on the ground. The impact of scintillation can be minimized by increasing the exposure time and maximizing the duty cycle, and through observing with larger telescopes at higher altitudes. Insofar as to allow for increased exposure times, diffusers do not specifically minimize scintillation noise. There have been suggestions in the literature on how to further minimize scintillation effects \citep{dravins1998,osborn2011}, including in particular, the conjugate plane photometry technique, which consists of putting a mask in the optical train at the conjugate plane of the scintillation layer to block out unwanted rays from the scintillation layer in the upper atmosphere. In doing so, the scintillation layer is effectively moved to the ground layers of the atmosphere, where the coherence angle is larger (on the order of $0.5^{\circ}$), allowing the intensity variations of the target star to be corrected by a comparison star. Although a promising method to suppress scintillation errors, and thus the photometric error budget as a whole, conjugate plane photometry requires a specific optical setup and specialized instrumentation. Furthermore, it also benefits from simultaneous SCIDAR measurements of the atmospheric turbulence to measure where the turbulent scintillation layer is and thus to inform where the conjugate mask should be placed \citep{osborn2011}. However, in our efforts to achieve precision photometry from the ground, we urge the community to consider such ways to suppress scintillation errors, and incorporating a diffuser with a scintillation-suppressing instrument, would potentially open a path to achieving even better precisions than presented here.

\vspace{0.5cm}

\begin{deluxetable*}{l l l l l l l l l l l}
\tabletypesize{\scriptsize}
\tablecaption{A comparison between the best ground-based photometric precision efforts in the literature to our knowledge. We choose to compare our photometric precision levels using two metrics: $\sigma_{\textrm{1min}}$, and $\sigma_{\textrm{30min}}$, where we assume that the former and latter metrics probe binning regimes weakly and strongly affected by correlate noise, respectively. Our highest precision efforts in the optical and NIR are shown in bold. For the $\sigma_{\textrm{30min}}$ metric, we only write the precision level value if we can find the value in the respective papers (e.g.,~from a precision vs bin plot). In the case where $\sigma_{\textrm{30min}}$ is well below the Gaussian expected value, we argue (see text) that that value is likely an overestimate of the actual precision achieved, and thus also write the Gaussian expected value (marked with a $\ddagger$). Overall, diffuser assisted photometry, both in the optical and the NIR, is achieving some of the best photometry in the literature. \label{tab:precision}}
\tablehead{\colhead{Band} & \colhead{$\sigma_{\textrm{unbinned}}$} & \colhead{$t_{\textrm{exp}}$} & \colhead{$t_{\textrm{dead time}}$} & \colhead{Duty Cycle} & \colhead{$\sigma_{\textrm{1min}}$}                          & \colhead{$\sigma_{\textrm{30min}}$}                          & \colhead{Method}              & \colhead{Diameter} & \colhead{Filter}        & \colhead{Reference}                 \\
	    \colhead{}                       & \colhead{[ppm]}                        & \colhead{[s]}                & \colhead{[s]}                      & \colhead{}           & \colhead{$\left[ \mathrm{\frac{ppm}{\sqrt{1min}}} \right]$} & \colhead{$\left[ \mathrm{\frac{ppm}{\sqrt{30min}}} \right]$} & \colhead{}                    & \colhead{[m]} & \colhead{}}
	    \startdata
	    Optical                          & 258                                    & 75                           & 30.5                               & 71.1\%               & 342.1                                                       & \nodata                                                      & Defocus                       & 4.0 & Corning 4-96        & \cite{gilliland1993}$^{\textrm{a}}$ \\
	                                     & 470                                    & 50                           & 29                                 & 63.3\%               & 533.7                                                       & $72^{26.3}_{-16}$$^*$                                            & Orthogonal-transfer CCD     & 2.2 & SDSS $z^\prime$         & \cite{johnson2009}$^{\textrm{b}}$   \\
	                                     & 603                                    & 8,11$^{\textrm{c}}$          & 35,32$^{\textrm{c}}$               & 35.7\%               & 455.7                                                       & $\sim$65$^\dagger$, $93^\ddagger$                            & Spectrophotometry & 10.4& Tunable Filter          & \cite{Colon2012}$^{\textrm{c}}$     \\
	                                     & 354                                    & 45                           & 0                                  & 100.0\%              & 306.6                                                       & $97^{33}_{-21}$$^*$                                              & Defocus                       & 3.5 & SDSS $r^{\prime}$       & \cite{kundurthy2013}$^{\textrm{d}}$ \\
	                                     & 258                                    & 120                          & 50                                 & 70.6\%               & 434.3                                                       & $87^{25}_{-17}$$^*$                                              & Defocus                       & 3.58& Gunn r                  & \cite{tregloan2013}$^{\textrm{e}}$  \\
	                                     & 211                                    & 150                          & 50                                 & 75.0\%               & 385.2                                                       & $124^{31}_{-22}$$^*$                                             & Defocus                       & 3.58& Gunn r                  & \cite{tregloan2013}$^{\textrm{e}}$  \\
	                                     & 910                       & 10                           & 4                                 & 71.4\%               & 423.6                                                       & $24^{14}_{-7}$$^\dagger$$^*$, $80^\ddagger$                                             & Defocus                       & 1.88& SDSS $r^\prime$          & \cite{fukui2016}$^{\textrm{f}}$  \\
	                                     & 1124                                   & 120                          & 11                                 & 91.6\%               & 1660.8                                                      & $246_{-81}^{+176}$                                           & Diffuser-assisted             & 0.6& Johnson I               & This work (55 Cnc)                 \\
	                                     & {\bf 494}                              & {\bf 16 }                    & {\bf 5 }                           & {\bf 76.2\% }        & {\bf 299.6 }                                                & {\bf $\mathbf{62^{+26}_{-16}}$ }                             & {\bf Diffuser-assisted}       & {\bf 3.5} & \textbf{Semrock 857/30} & {\bf This work (16 Cygni)}           \\
	                                     & 1771                                   & 6                            & 2.5                                & 80.0\%               & 626.1                                                       & $180_{-41}^{+66}$                                            & Diffuser-assisted             & 3.5& SDSS $r^\prime$         & This work (WASP 85 A)                \\
	                                     & 750                                    & 30                           & 2.5                                & 92.3\%               & 541.0                                                       & $54_{-14}^{+23}$$^\dagger$,101$^\ddagger$                    & Diffuser-assisted             & 3.5& SDSS $i^\prime$         & This work (TRES-3b)                  \\ \hline
	    NIR                              & 1780                                   & 5                            & 9.7                                & 34\%                 & 860                                                         & $\sim$200$^\dagger$                                          & Defocus                       & 3.6& $K_S$                   & \cite{croll2011}$^{\textrm{g}}$      \\
	                                     & 5103                                   & 8                            & 15.5                               & 34.0\%               & 3195.6                                                      & $\sim$406$^\dagger$                                          & Defocus                       & 5.0& $K_S$                   & \cite{zhao2014}$^{\textrm{h}}$       \\
	                                     & {\bf 1232 }                            & \textbf{40}                  & {\bf 15 }                          & {\bf 72.7\%   }      & {\bf 1182.4}                                                & $\mathbf{137^{+64}_{-36}}$                                      & {\bf Diffuser-assisted }      & {\bf 5.0} & $\mathbf{K_S}$          & {\bf This work }                     \\
\enddata
\tablenotetext{*}{Calculated from the published light-curve data-points, calculating residuals vs. binning through fitting a best-fit transit model (in a similar fashion as for the diffuser-assisted observations).}
\tablenotetext{\dagger}{Estimated from precision vs binning plots in the respective papers.}
\tablenotetext{\ddagger}{The values are the Gaussian expected precision level in 30 minutes, and are likely closer to the truer precision achieved in these cases.}
\tablenotetext{a}{Using the 4m telescope at Kitt Peak National Observatory, using a Corning glass filter, with a central wavelength of 472nm, and a FWHM of 166.5nm, observing stars in the M67 cluster.}
\tablenotetext{b}{Using the Orthogonal Parallel Transfer Imaging Camera on the 2.2m University of Hawaii telescope at Maunakea, observing the transit of WASP 10 b.}
\tablenotetext{c}{Using the narrow-band ($\sim$1.2nm) tunable filter capability of the OSIRIS instrument on the 10.4m GTC centered around the KI line (769.75nm), observing in-and-out-of transit of HD80606b. The observations in \cite{Colon2012} were staggered observations in different band-passes. The observations we picked to include are the out-of-transit observations, which gave the best precision at long binning timescales. These observations had a fixed 4s filter-tuning time, followed by an observation in another band. The exposure times were switched between 8s and 11s, and thus a varying dead time between. We downloaded the supplementary data from the paper (Table 5), and calculated the mean time between successive exposures, which was 43s in the highest cadence band (770nm band). This band was the highest cadence and the highest precision at the longest bin sizes.}
\tablenotetext{d}{Using AGILE, a charge transfer CCD enabling 100\% duty cycle \citep{mukadam2011,kundurthy2011}, on the 3.5m telescope at Apache Point Observatory, observing the transit of XO-2b.}
\tablenotetext{e}{Using the 3.58m ESO New Technology Telescope at La Silla, Chile, using the Gunn $r$ filter (ESO Filter \#784) to observe the transit of WASP 50 b.}
\tablenotetext{f}{Using the 1.88m telescope at the Okayama Astrophysical Observatory in Japan, to observe the transit of HAT-P-14b in multiple bands with the MuSCAT imager (SDSS $r^\prime$ were the highest precision observations).}
\tablenotetext{g}{Using the 3.6m Canada-France-Hawaii Telescope (CFHT) at Maunakea, to observe the secondary eclipse of WASP 12 b.}
\tablenotetext{h}{Using the 5.0m Hale 200" telescope at Palomar, to observe the secondary eclipse of HAT-P-32Ab.}
\end{deluxetable*}

\subsection{Adaptability of diffusers in other systems}
\label{sec:grating}
The most straight-forward way to incorporating a diffuser in a telescope imaging system is in a standard filter wheel. This is most efficient if the telescope has two filter wheels, placing the diffuser in one slot of one filter wheel, retaining the capability to select a filter in the other wheel. If only one filter wheel is available, a diffuser could be combined with a filter in a single filter wheel slot by placing them back-to-back in the slot. This configuration could potentially cause back-reflections and ghosting, which could be corrected by adding a small wedge between the filter and a diffuser. A more permanent solution would be imprinting a diffuser pattern directly onto the filter. Although we have not developed such a device, it is an interesting avenue for further study.

We studied the PSFs of off-the-shelf top-hat diffusers. The off-the-shelf diffusers studied are capable of spreading out the light deterministically over a large number of pixels, which can easily be calculated using Equation \ref{eq:fwhm}, or through precisely modeling diffusers using Zemax OpticStudio. Although all of the off-the-shelf diffusers tested produce an approximate a top-hat shaped PSF, we observed them to have a speckle pattern of $\sim$20-40\% of the total intensity along with having less than optimal wings. Although the speckle pattern is observed to be completely stable in the lab and on sky, we demonstrate that both the speckle pattern and the wing fall-off can be further reduced through optimizing the design of the diffuser for a given application. This optimization process improves the signal within the aperture, but we demonstrate that using off-the-shelf diffusers can yield very high-precision photometry even on small telescopes.

Rotation is effective at removing the speckles observed on the diffuser PSF. Although we found that even without rotation the diffuser PSF is extremely stable, rotation helps further smooth out the PSF. Rotational smoothing effectively increases the dynamic range of the observations, removing any spikes or speckles that might potentially saturate the detector. 

Diffusers, however, are not free of limitations. Their main drawback is their fixed PSF size. In crowded fields, such as fields toward the center of the galaxy sources can overlap. This can similarly be an issue for closely separated binary stars, but at least in the case of our WASP 85 A b observations, we were able to demonstrate the capability to maintain a high level of precision throughout the observations, even with completely overlapping PSFs. Another limitation is background sky-noise, which starts to dominate for stellar PSFs spread out over many pixels, an especially important consideration for faint stars, and in the NIR. In practice, for our diffuser-assisted observations with ARCTIC, we start to see significant effects from background sky noise at around SDSS $i^\prime$ magnitudes of $\sim$13-14. One way to solve this problem and enable adjustability in the PSF size involves varying the distance of the diffuser from the detector to deterministically optimize the size of the PSF for a given observation. This, however, would require a dedicated pistoning mechanism that could move the diffuser towards and away from the detector, which might not always be possible. Another option would be to have a separate filter wheel with diffusers of different opening angles. The price of such a setup could be kept to a minimum by using off-the-shelf polymer diffusers costing $\sim$\$250 each.

An additional second-order effect that we have observed with our final optical diffuser on ARCTIC is that around very bright stars, the diffuser forms images of two equally bright spots equally distant from the main PSF $\sim$800 pixels away from the central PSF, that have an intensity proportional to the net counts in the central PSF. We note that these spots are faint: the total counts observed in the diffractive orders is $\sim$$10^4$ counts for $4\times10^7$ counts in the central PSF. We speculate that the spots observed are the +1 and -1 orders of the diffuser acting as a diffraction grating, as we observe that the exact separation of the spots is wavelength dependent. Through modeling the diffuser in the telescope as a transmissive diffraction grating in Zemax, we can accurately replicate the spot locations, assuming a groove spacing of 6.0 microns. We attribute this grating effect to potential periodic systematics in the laser writer during the diffuser fabrication process, where 6 microns might correspond to the row step size, or an integer multiple of the step size. We stress that the effect has little-to-no effect on the photometric precision, as it is a static effect. If, however, the spot was to completely overlap one of the reference stars, it would introduce an additional source of systematic error, correlated to the signal being studied. This can be easily solved by rotating the diffuser, or by rotating the focal plane array with respect to the diffuser to make sure that the ghost spots from one star do not coincide with a target or a reference star in the field.

\subsection{Using diffusers in space?}
Engineered Diffusers\texttrademark~are available with direct etching of the diffusing surface on fused silica, making it suitable for space applications due to its radiation resistant properties. As such, Engineered Diffusers\texttrademark~have been considered in the lab for space use for the CHEOPS mission by \cite{magrin2014}. Their aim was to shape the PSF to a nearly top-hat with 30" FWHM, corresponding to a 30pixel FWHM due to the CHEOPS plate scale of 1".0/pixel. 

\cite{magrin2014} studied in the lab the PSF shaping capabilities an off-the-shelf Engineered Diffuser\texttrademark~(part number: EDC-0.25-A-1r), with an opening angle of $0.25^{\circ}$. This is the same $0.25^{\circ}$ polymer diffuser pattern as we studied (see Table \ref{tab:diffusers}), replicated on a 1" circular substrate instead of the 2"x2" square substrate diffuser we studied. \cite{magrin2014} also studied using a microlens array from RPC Photonics to shape the PSF to the desired top-hat form factor. In their study, \cite{magrin2014} found that the resulting diffuser PSF shape did not approximate a top-hat shape, but rather had broad Gaussian-like output with numerous speckles. This is similar to our results with the $0.25^{\circ}$ diffuser, as we show in Figure \ref{fig:rotationpanel1}: the $0.25^{\circ}$ diffuser indeed gave the largest amplitude speckles of the off-the-shelf diffusers tested. However, in this paper we have demonstrated that the speckling is completely stable throughout the observations, and that both the speckling and the fall-off of the wings can be further optimized to deliver a homogeneous PSF shape approximating a top-hat shape. The CHEOPS team decided to not fly with a diffuser choosing to defocus the telescope instead, due to the spikiness of the observed diffused PSF and the additional risk associated with what to them was unproven technology for high-precision photometry applications.

We now consider diffusers a proven technology in achieving high-precision from the ground. We believe that in space too, diffusers will be beneficial for high photometric precision surveys of bright nearby targets---especially with a diffuser customized for the telescope system. Diffusers minimize flat-fielding errors, and jitter effects in the pointing of the telescope, a major source of systematics in space telescope systems. We speculate that diffusers could enable small spacecraft, such as cubesats, with less-than-optimal pointing precision to achieve high-precision photometry. This could be especially beneficial e.g.,~to perform long-term uninterrupted photometric monitoring of RV planet hosts, to separate stellar activity from exoplanet signals.

\subsection{Future outlook}
Our results highlight the potential of using ground-based diffuser-assisted photometry to perform routine high-precision follow-up observations of nearby bright planet-systems, such as those that are being detected with \textit{K2} and those that will be detected by the TESS mission in the future. Diffuser-assisted photometry systems from the ground will allow for consistent rapid reconnaissance follow-up observations for TESS targets, and through spreading out diffusers to other telescopes, we expect that telescopes equipped with diffusers will serve as work-horse instruments in following up with TESS candidates in the future. Through this effort, we can secure an up-to-date target list of precisely characterized planets for further study of exoplanet atmospheres with facilities such as Hubble, JWST, and future 30m-class telescopes.

In the interest of spreading this effort and technology to the community for use on other telescopes, our intent is to make our optimized diffuser pattern for the optical diffuser available to the community through RPC Photonics.

\section{Summary}
We describe a reliable technique to achieve space-quality photometric precision on nearby bright stars using ground-based telescopes, by coupling a beam shaping diffuser capable of molding a varying stellar input to a broad stabilized top-hat shape without defocusing the telescope. Spreading the light over many pixels minimizes flat field errors and telescope guiding errors, allowing exposure times to be increased to effectively gather more photons while averaging over scintillation errors. Using this technology, we have demonstrated some of the highest photometric precisions from the ground on the ARCTIC camera on the Apache Point 3.5m telescope in the optical, and the WIRC camera at 200" Hale telescope at Palomar in the NIR. Specifically, on ARCTIC we achieve $62^{+26}_{-16}$ppm precision in 30 minute bins on a nearby bright star 16 Cyg A, and with WIRC we achieve $137^{+64}_{-36}$ppm precision in an early test demonstration of diffusers in the NIR. Additionally, we demonstrate that diffuser-assisted observations on small type telescopes are also capable of delivering precision photometry ($\sim$300ppm in 30 minutes) in observations using the Penn State PlaneWave CDK 24" telescope of 55 Cnc.

In this paper we have discussed how diffusers operate, reporting our lab and on-sky tests with different telescope systems, demonstrating that diffusers offer broad band compatibility, in both the optical and NIR. Moreover, through both numerical simulations and on-sky efforts, we demonstrate that diffusers work in both collimated and converging telescope beams. Being relatively simple and inexpensive devices, diffusers can be easily incorporated into a variety of telescope systems to improve their photometric precision on nearby bright stars. The true power of this technique is making high photometric precision levels widely accessible without specialized instrumentation, or the rarest observing conditions. In the light of the upcoming TESS mission, which will require precise and timely follow-up from the ground to validate and characterize exciting transiting targets, our intent is to work with the community to spread out diffuser technology by offering our optimized diffuser pattern through RPC Photonics.

\acknowledgments
We thank the anonymous referee for a thoughtful reading of the manuscript, and for useful suggestions and comments. We gratefully acknowledge the work and assistance of Tasso Sales and Laura Weller-Brophy at RPC Photonics, without whose help this project would not have been possible. GKS wishes to thank Eric Ford for helpful discussions on MCMC fitting and inference. This work was directly seeded and supported by a Scialog grant from the Research Corporation for Science Advancement (Rescorp) to SM, LH, JW. This work was partially supported by funding from the Center for Exoplanets and Habitable Worlds. The Center for Exoplanets and Habitable Worlds is supported by the Pennsylvania State University, the Eberly College of Science, and the Pennsylvania Space Grant Consortium. GKS acknowledges support from the Leifur Eiriksson Foundation Scholarship. This work was supported by NASA Headquarters under the NASA Earth and Space Science Fellowship Program-Grant NNX16AO28H. This work was performed in part under contract with the Jet Propulsion Laboratory (JPL) funded by NASA through the Sagan Fellowship Program executed by the NASA Exoplanet Science Institute. We acknowledge support from NSF grants AST-1006676, AST-1126413, AST-1310885, AST-1517592, the NASA Astrobiology Institute (NAI; NNA09DA76A), and PSARC.

These results are based on observations obtained with the Apache Point Observatory 3.5-meter telescope which is owned and operated by the Astrophysical Research Consortium, the Hale 200" Telescope at Palomar Observatory, and the Planewave CDK24 Telescope operated by the Penn State Department of Astronomy \& Astrophysics at Davey Lab Observatory. The Palomar Hale 200 inch telescope is operated by Caltech and the Jet Propulsion Laboratory. This paper includes data collected by the \textit{Kepler} telescope. The \textit{Kepler} and \textit{K2} data presented in this paper were obtained from the Mikulski Archive for Space Telescopes (MAST). Space Telescope Science Institute is operated by the Association of Universities for Research in Astronomy, Inc., under NASA contract NAS5-26555. Support for MAST for non-HST data is provided by the NASA Office of Space Science via grant NNX09AF08G and by other grants and contracts. Funding for the \textit{K2} Mission is provided by the NASA Science Mission directorate. This research made use of the NASA Exoplanet Archive, which is operated by the California Institute of Technology, under contract with the National Aeronautics and Space Administration under the Exoplanet Exploration Program.

Facility: ARC 3.5m, Palomar 200", Penn State PlaneWave CDK 24", \textit{Kepler}, \textit{K2}.

Software:  AstroImageJ \citep{collins2017}, \texttt{astropy, astroscrappy, BATMAN} \citep{kreidberg2015}, \texttt{corner.py} \citep{dfm2016}, \texttt{emcee} \citep{dfm2013}, Everest 2.0 \citep{luger2017}, MC3 \citep{cubillos2017}, TERRASPEC \citep{bender2012}, Zemax OpticStudio.


\pagebreak
\bibliographystyle{yahapj}
\bibliography{references}

\appendix

\section{\textit{Kepler} and \textit{K2} photometry}
\begin{figure*}[h]
	\begin{center}
		\includegraphics[width=0.8\textwidth]{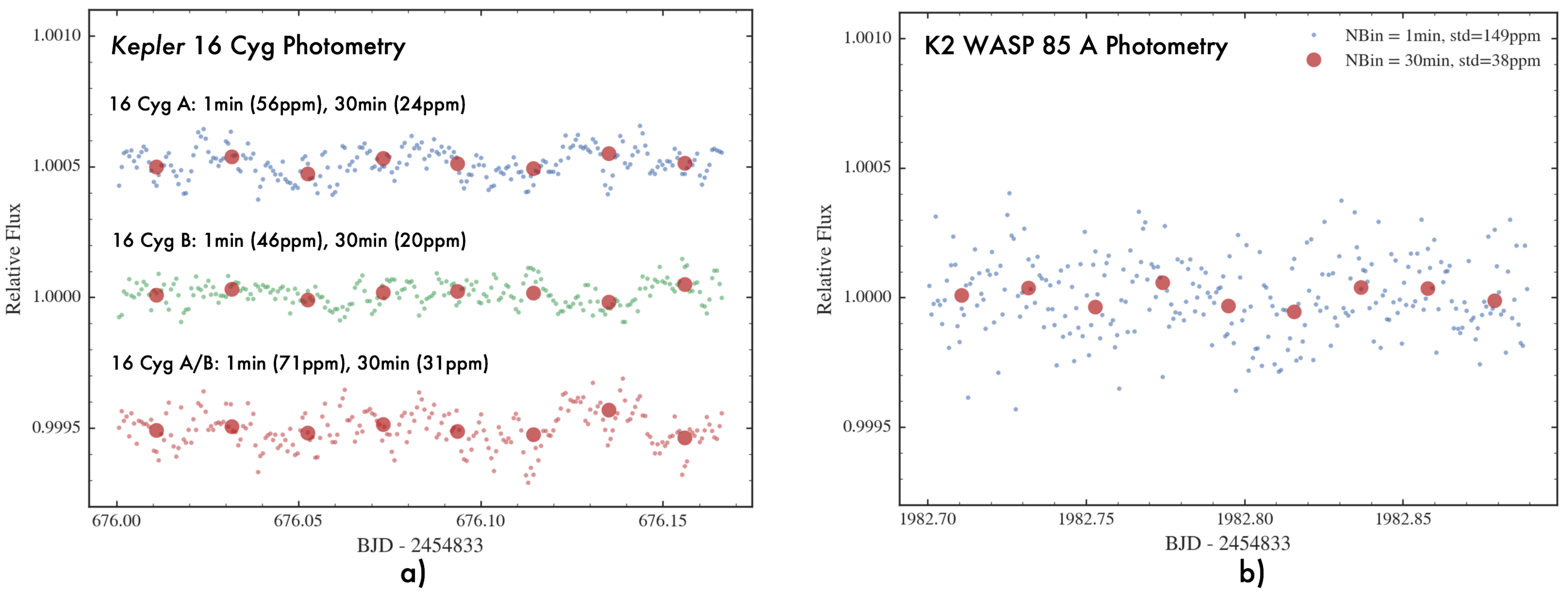}
	\end{center}
    \vspace{-0.5cm}
	\caption{Detrended \textit{Kepler} and \textit{K2} short cadence photometry of the 16 Cyg system (a), and WASP 85 A (b).  The data are plotted on the same scale. a) 4hours of \textit{Kepler} short cadence data of 16 Cyg A used in Figure \ref{fig:16cygnipanel}. Also shown is the same period for 16 Cyg B, along with the \textit{Kepler} differential light curve for 16 Cyg A using 16 Cyg B as a reference star (light curve of 16 Cyg A divided by the light curve of 16 Cyg B). b) \textit{K2} data for WASP 85 A used in Figure \ref{fig:wasp85panel}. These data were retrieved from MAST, and the Everest 2.0 website, for panels a and b, respectively.}
	\label{fig:kepler_panel}
\end{figure*}

Figure \ref{fig:kepler_panel}a shows the section of 4 hour short cadence photometry of 16 Cyg A (blue curve, unbinned) and 16 Cyg B (green curve, unbinned), as observed by \textit{Kepler} in Quarter 7, used in Figure \ref{fig:16cygnipanel}. Additionally shown in Figure \ref{fig:kepler_panel}a is the differential photometry of 16 Cyg A, using 16 Cyg B as a reference star (red curve, unbinned, denoted 16 Cyg A/B). Furthermore, we also show the 30 minute binned points for each curve (thick red dots), along with the unbinned and binned precision for this segment for the three light curves. We chose this specific time window to be representative of the \textit{Kepler} data of 16 Cyg A, as in 30 minute bins, the RMS scatter of this 4 hour segment matches well with the 30 minute CDPP precision across the whole Quarter 7 of this star. These data were easily retrievable from MAST, and detrended as described in the text.

Similarly, figure \ref{fig:kepler_panel}b shows the section of 4.5 hour short cadence photometry of WASP 85 A, as observed by \textit{K2} in Campaign 1, used in Figure \ref{fig:wasp85panel}. The short-cadence detrended \textit{K2} data of WASP 85 A was retrieved and detrended using the Everest 2.0 website and pipeline\footnote{Everest 2.0 website: \url{http://staff.washington.edu/rodluger/everest/catalog.html}}, but the data is also readily available from MAST.

We note that we can visibly see evidence of correlated noise structure in both the 16 Cyg and the WASP 85 A photometry, which we attribute to astrophysical activity. The scatter for WASP 85 A is visibly larger than that for the 16 Cyg system due to its faintness.

\newpage

\section{Transit fitting posterior plots}
\subsection{Transit of WASP 85 A b}
\begin{figure*}[h!]
	\begin{center}
		\includegraphics[width=\textwidth]{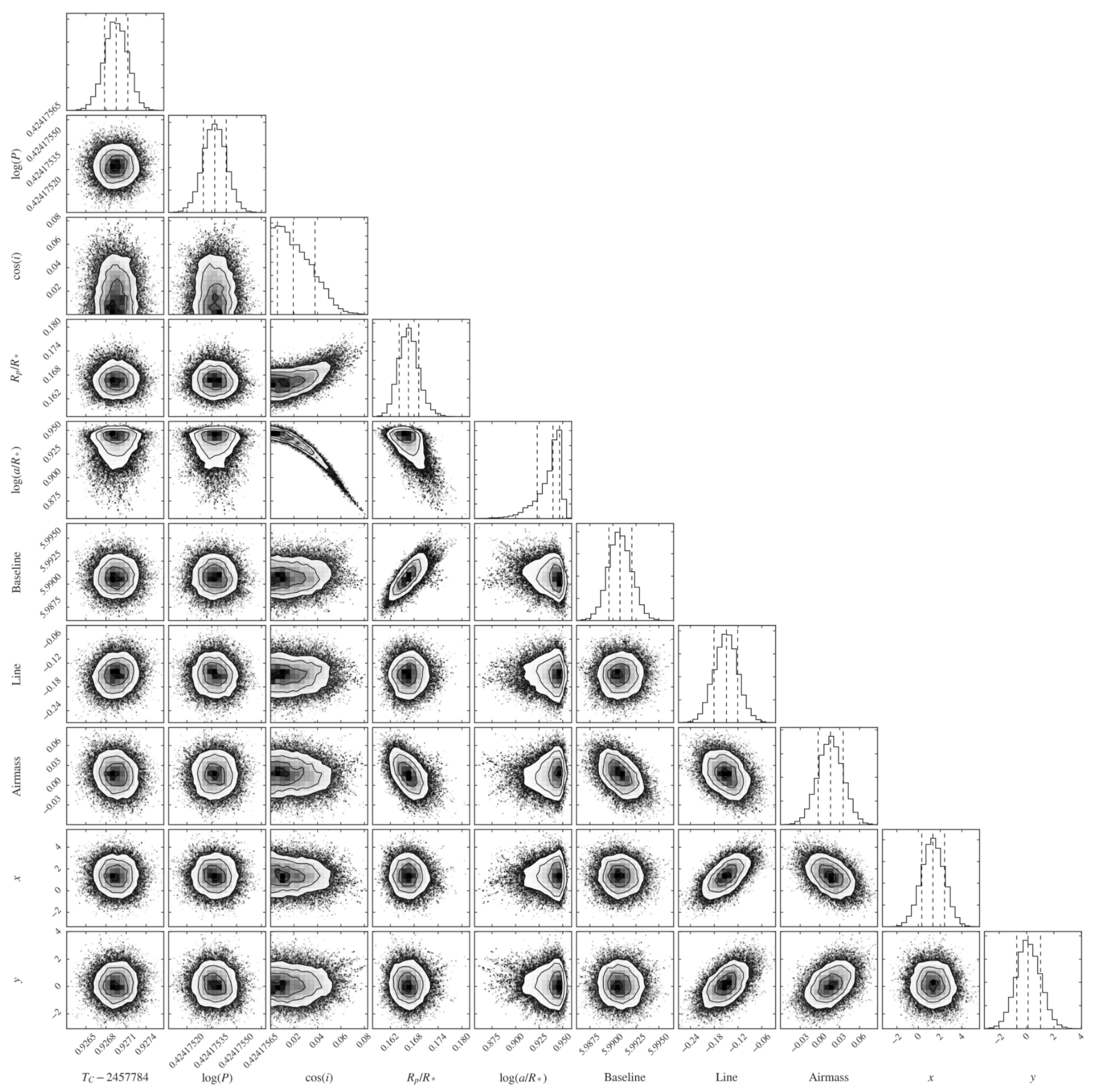}
	\end{center}
	\caption{Corner plot from WASP 85 A b data. Plot created using \texttt{corner.py} \citep{dfm2016}. }
	\label{fig:wasp_corner}
\end{figure*}

\newpage

\subsection{Transit of TRES-3b}

\begin{figure*}[h!]
	\begin{center}
		\includegraphics[width=\textwidth]{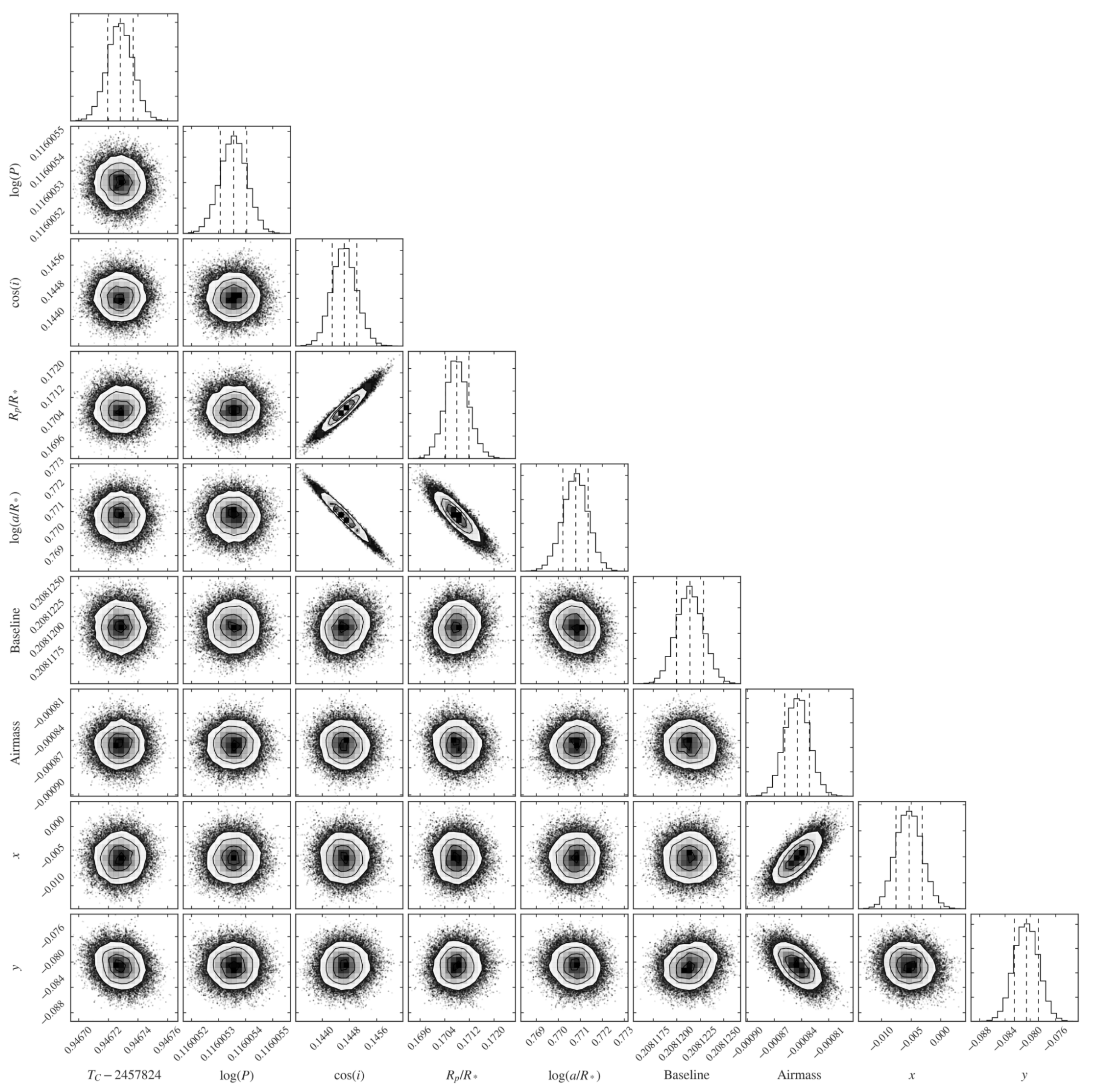}
	\end{center}
	\caption{Corner plot from TRES-3b data. Plot created using \texttt{corner.py} \citep{dfm2016}. }
	\label{fig:tres_corner}
\end{figure*}

\end{document}